\documentclass[lettersize,journal]{IEEEtran}
\usepackage{amsmath,amsfonts}
\usepackage{algorithmic}
\usepackage{algorithm}
\usepackage{array}
\usepackage[caption=false,font=normalsize,labelfont=sf,textfont=sf]{subfig}
\usepackage{textcomp}
\usepackage{stfloats}
\usepackage{float}
\usepackage{url}
\usepackage{verbatim}
\usepackage{graphicx}
\usepackage{cite}
\usepackage{siunitx}
\usepackage{balance}
\usepackage[export]{adjustbox}
\usepackage{circuitikz}
\usepackage{color,soul}
\begin{document}
\bstctlcite{IEEEexample:BSTcontrol}

\title{A Distributed Magnetostatic Resonator}

\author{Connor Devitt,~\IEEEmembership{Graduate Student Member,~IEEE}, Sudhanshu Tiwari,~\IEEEmembership{Member,~IEEE}, \\Sunil A. Bhave,~\IEEEmembership{Senior Member,~IEEE}, Renyuan Wang,~\IEEEmembership{Member,~IEEE}
	% <-this % stops a space
	\thanks{Manuscript received on XX XX, 2023; revised on XX XX, 2024; accepted on XX XX, 2024. This research was developed with funding from the Air Force Research Laboratory (AFRL) and the Defense Advanced Research Projects Agency (DARPA). The views, opinions and/or findings expressed are those of the authors and should not be interpreted as representing the official views or policies of the Department of Defense or the U.S. Government. This manuscript is approved for public release; distribution A: distribution unlimited. (\textit{Corresponding authors: Connor Devitt, Sunil A. Bhave})}	
    \thanks{C.D. and S.T. designed the resonators. RW designed the patterned plating process. Chip fabrication, resonator measurement, and filter simulations were performed by C.D. Manuscript was prepared by C.D. with inputs from S.T., S.A.B., and R.W.}
	\thanks{Connor Devitt (e-mail: devitt@purdue.edu), Sudhanshu Tiwari (e-mail: tiwari40@purdue.edu), and Sunil A. Bhave (e-mail: bhave@purdue.edu) are with the OxideMEMS Lab, Elmore Family School of Electrical and Computer Engineering, Purdue University, West Lafayette, IN 47907 USA.}% <-this % stops a space
	\thanks{Renyuan Wang is with FAST Labs, BAE Systems, Inc., Nashua,
		NH 03060 USA (e-mail: renyuan.wang@baesystems.com).}}

%\markboth{Transactions on Microwave Theory and Techniques}
{}

\maketitle

\begin{abstract}
This work reports the design, fabrication, and characterization of coupling-enhanced magnetostatic forward volume wave resonators with significant spur suppression. The fabrication is based on surface micro-machining of yttrium iron garnet (YIG) film on a gadolinium gallium garnet (GGG) substrate with thick gold transducers. A distributed resonator is used to excite forward volume waves in YIG to realize a frequency dependent coupling boost. Fabricated devices at 18 GHz and 7 GHz show coupling coefficients as high as 13\% and quality factors above 1000. Higher-order magnetostatic mode suppression is experimentally demonstrated through a combination of transducer and YIG geometry design. An edge-coupling filter topology is proposed and simulated which utilizes this novel distributed magnetostatic resonator.
\end{abstract}


\section{Introduction}
\IEEEPARstart{N}{ext} generation RF front-end modules require low-loss and highly-selective bandpass filters to cover the multitude of densely allocated frequency bands for commercial communication systems scaling beyond $\SI{6}{\giga\hertz}$. Highly compact bandpass filters are also desired for wideband active electronically steered antenna arrays (AESA), which are susceptible to interference and jamming when beamforming electronics are integrated at the element level. Including either tunable or arrays of selective filters between the antenna and receiver electronics mitigates this risk while maintaining the advantages of fully-digital wideband AESAs \cite{ariturk_element-level_2022,cdi_proquest_reports_2564179163, talisa_benefits_2016}. Although significant progress has been made with a variety of electromagnetic filter technologies \cite{snyder_emerging_2021} in terms of synthesis, design, and manufacturing techniques, the filter size is fundamentally limited by the electromagnetic wavelength. Acoustic filters have seen significant commercial success for sub-$\SI{6}{\giga\hertz}$ filtering due to their high quality factors (Q-factors), scalable fabrication, and compact size orders of magnitude smaller than their electromagnetic counterparts \cite{aigner_saw_2008}. Scaling acoustic resonators towards higher frequency while maintaining high Q and large coupling coefficients to realize low loss, wide bandwidth filters remains an active area of research \cite{hagelauer_microwave_2023,yang_scaling_2018, barrera_thin-film_2023, fiagbenu_k-band_nodate, zou_aluminum_2022, giribaldi_compact_2023}. 

\begin{figure}[t]
	% \vspace{-0.1in}
	\begin{center} 
		\noindent
		\subfloat[]{\includegraphics[width=2.8in]{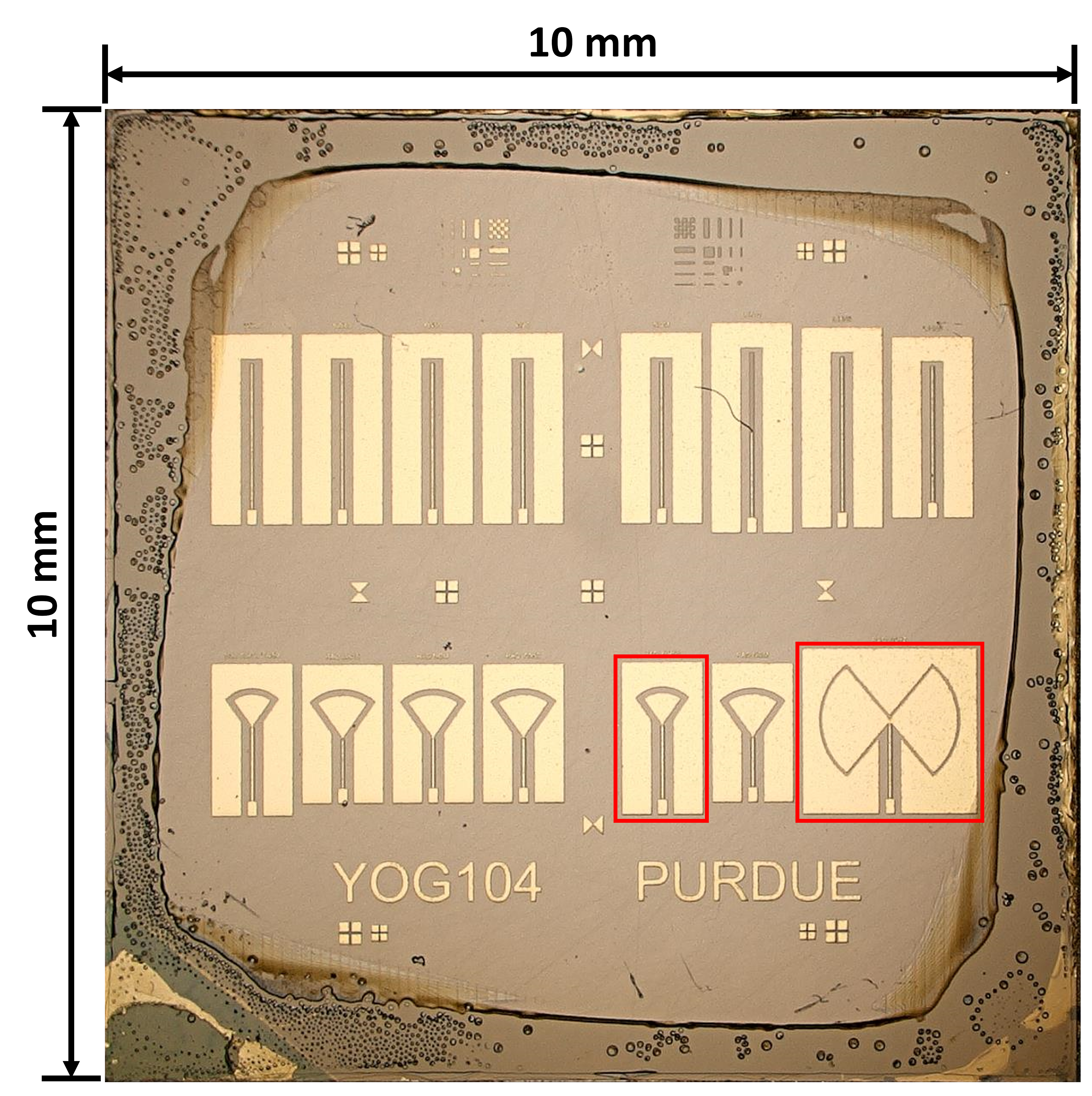}%
			\label{whole_chip}}
		\hfil
		\subfloat[]{\includegraphics[width=2.8in]{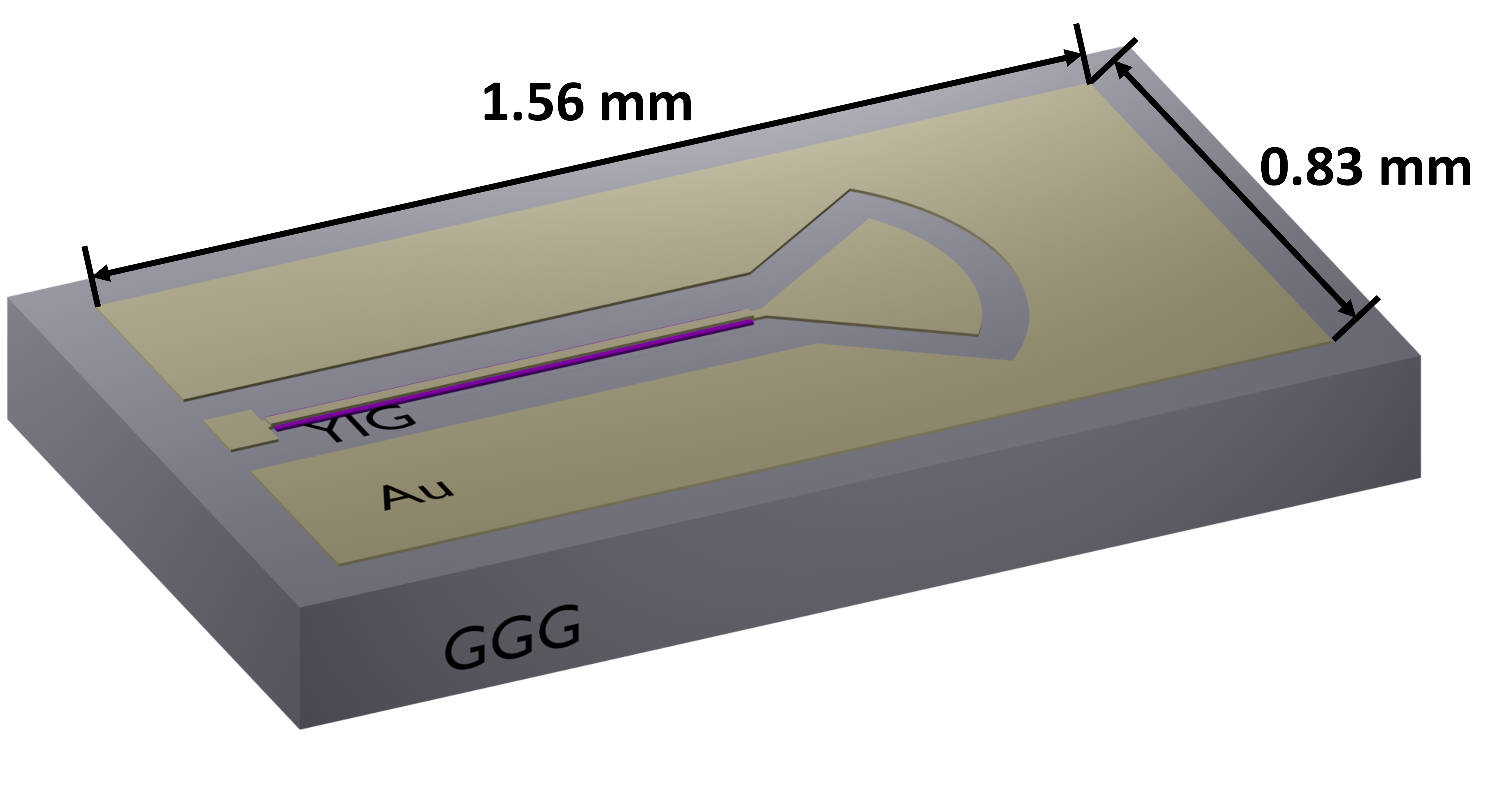}%
			\label{chip_cartoon}}
		%	\hfil
		
		\caption{\textbf{(a)} Microphotograph of multiple magnetostatic forward volume wave resonators fabricated on a YIG on GGG chip. Highlighted in red are two resonators designed at \SI{18}{\giga\hertz} (left) and \SI{7}{\giga\hertz} (right). \textbf{(b)} Rendering of the distributed resonator.}
		\label{chip_pictures}
	\end{center}
	\vspace{-0.3in}
\end{figure}

Filters based on magnetic materials such as yttrium iron garnet (YIG) have been an attractive technology for many years due to the material's magnetically tunable dispersion relation and low Gilbert damping. The geometry of a magnetostatic wave (MSW) cavity can be designed independently from its bias-dependent resonant frequency allowing significant miniaturization over electromagnetic resonators. However, this technology has faced a number of integration and scaling challenges. Until recently MSW filters were constructed out of flip-chip assemblies on a PCB \cite{adam_msw_1985, hanna_single_1988, yang_low-loss_2013, zhang_nonreciprocal_2020} or using polished YIG spheres manually aligned to 3D transducers limiting the device scalability and cost \cite{noauthor_yig_nodate, fletcher_ferrimagnetic_1959, carter_magnetically-tunable_1961, ishak_magnetostatic_1988}. The YIG needs to be biased with a strong magnetic field using either a rare-earth permanent magnet or a high-power electromagnet \cite{du_frequency_2023, noauthor_yig_nodate,tikhonov_temperature_2013}, increasing the system size and complexity despite the compact filter core. Additionally, previously reported magnetostatic coupling coefficients (analogous to the acoustic piezoelectric effective coupling coefficient) have been small ($<3\%$) \cite{devitt, du_frequency_2023, dai_octave-tunable_2020} constraining the maximum filter bandwidth and limiting potential applications. Shorted microstrip transducers \cite{marcelli_band-pass_2004} can give relatively high coupling, but the resonator dimensions become moderately large. Recently, a low-power integrated tunable biasing method has been demonstrated in \cite{du_frequency_2023}, allowing system miniaturization. Scalable lithographic fabrication methods have also been demonstrated using wet etching \cite{du_frequency_2023}, reactive ion etching (RIE) \cite{costa_compact_2021,connelly2023principles}, and physical etching \cite{devitt, feng_micromachined_2023, dai_octave-tunable_2020}. In this work, a new compact distributed resonator design is presented (Fig. $\ref{chip_pictures}$) which enables wider bandwidth filters at the expense of magnetic tuning range.

\section{Resonator Design and Modeling}

\begin{figure}[t]
	\centering
	\subfloat[]{
		\begin{circuitikz}[scale=0.8]
			\ctikzset{label/align=rotate}
			
			\ctikzset{capacitors/scale=0.4}
			\ctikzset{resistors/scale=0.5}
			\ctikzset{inductors/scale=0.65}
			
			\draw (7,0) to (7,0) node[ground]{}; 
			
			\draw
			(0,0)  node[label={[font=\footnotesize]left:$P_{in}$}] {}
			(2,0) to[R=$R_0$, -] (0,0)
			(2,0) to[L= $L_0$] (4,0)
			%		(3,0)  node[label={[font=\footnotesize]below:$P_0$}] {}
			
			(4,1) to[R=$R_{m1}$] (6,1)
			(4,0) to[L=$L_{m1}$] (6,0)
			(4,-1) to[C=$C_{m1}$] (6,-1)
			(6,0) to[open, -] (7,0)
			
			(4,1) -- (4,-1)
			(6,1) -- (6,-1)
			(6,0)-- (7,0)
			;
		\end{circuitikz}
		\label{lumped_circuit}}
	\vspace*{-0.15in}
	\\
	\subfloat[]{
		\begin{circuitikz}[scale=0.8]
			\ctikzset{label/align=rotate}
			
			\ctikzset{capacitors/scale=0.4}
			\ctikzset{resistors/scale=0.5}
			\ctikzset{inductors/scale=0.65}

                \draw [black,domain=-45:45] plot ({0.5*cos(\x)+6.5 - 0.5*cos(45)}, {0.5*sin(\x)});
                \draw[black](6,0) -- (6.5,{0.25*sqrt(2)});
                \draw[black](6,0) -- (6.5,{-0.25*sqrt(2)});
                
			\draw
			(0,0)  node[label={[font=\footnotesize]left:$P_{in}$}] {}
			(2,0) to[R=$R_0$, -] (0,0)
			(4,0) to[generic= $Z_0${$,\,$}$P_0$] (6,0)
			%		(3,0)  node[label={[font=\footnotesize]below:$P_0$}] {}
			
			(2,1) to[R=$R_{m1}$] (4,1)
			(2,0) to[L=$L_{m1}$] (4,0)
			(2,-1) to[C=$C_{m1}$] (4,-1)
			% (6,0) to[open, -o] (7,0)
			
			(4,1) -- (4,-1)
			(2,1) -- (2,-1)
			% (6,0)-- (7,0)
			;
		\end{circuitikz}
		\label{distributed_circuit}}
	\caption{\textbf{(a)} Conventional lumped element model for electrically short MSW resonators with shorted termination. \textbf{(b)} Distributed model for MSW resonators terminated with a high-frequency short where $Z_0$ and $P_0$ represent the characteristic impedance and physical length of a transmission line.}
	\label{circuit_models}
	\vspace*{-0.2in}
\end{figure}
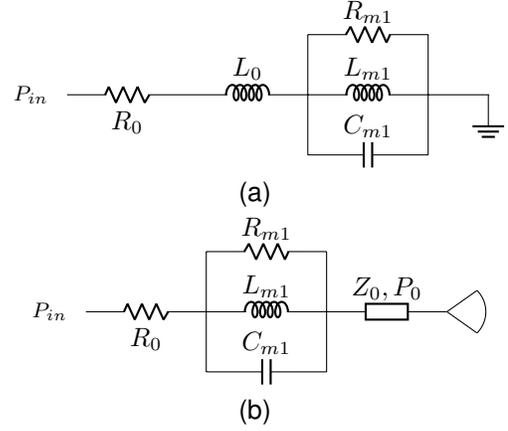

In previous work, a MSW is excited in YIG using an inductive element with a frequency response modeled using the lumped circuit shown in Fig. $\ref{lumped_circuit}$ as reported in \cite{dai_octave-tunable_2020,feng_micromachined_2023, gao_design_2022, gao_equivalent_2021, cui_coupling_2019, devitt}. $R_0$ and $L_0$ represent the equivalent loss and inductance in the microwave transducer while $R_m$, $C_m$, and $L_m$ represent a single resonant MSW mode at a frequency $f_m$. The resonator exhibits a MSW resonance and anti-resonance for each excited mode whose frequency can be calculated from the YIG geometry using the dispersion relation for magnetostatic forward volumes waves (MSFVW) \cite{stancil_theory_1993}

\begin{equation}
	\omega^2 = \omega_0\left[\omega_0+\omega_m\left(1 - \frac{1-e^{-k_{mn}t}}{k_{mn}t}\right)\right],
	\label{msfvw_dispersion}
\end{equation}
where $\omega_m = \mu_0\gamma_m M_s$, $\omega_0 =\mu_0\gamma_m H^{eff}_{DC}$, $t$ is the film thickness, $\gamma_m$ is the gyromagnetic ratio, $\mu_0$ is the permeability of free space, $k_{mn}$ is the wave vector, $M_s$ is the saturation magnetization, and $H^{eff}_{DC}$ is the effective DC magnetic bias. The MSW is confined within the YIG mesa forming a number standing waves with discrete wave vectors \cite{ishak_tunable_1986, ishak_tunable_1988, hanna_single_1988,devitt}. Analogous to acoustic resonators, a magnetostatic effective coupling coefficient can be defined as
\begin{equation}
    k_{eff}^2 = \frac{\pi}{2}\left(\frac{f_1}{f_2}\right)\cot\left(\frac{\pi}{2}\frac{f_1}{f_2}\right)
    \label{coupling}
\end{equation}

For the shorted transducers (Fig. $\ref{lumped_circuit}$), $f_1$ is the MSW resonance ($f_m$) and $f_2$ is the anti-resonance. With this model, $k_{eff}^2$ is a function of the ratio of $L_m$ to $L_0$ and limits the maximum achievable filter bandwidth for a specified level of passband ripple. For acoustic resonators modeled using the Butterworth-Van Dyke circuit \cite{larson_modified_2000}, $k_{eff}^2$ (related to the ratio motional capacitance $C_m$ to static capacitance $C_0$) also limits the maximum bandwidth. Several groups \cite{psychogiou_hybrid_2015, yang_x-band_2021, truitt_efficient_2007, cagdaser_low-voltage_2012,barzanjeh_mechanical_2017,yang_modified_2016, xie_sub-terahertz_2023} have demonstrated micro-mechanical and bulk acoustic resonators with enhanced $k_{eff}^2$ beyond the material limited electromechanical coupling using inductors in parallel with the acoustic resonator to compensate for $C_0$. However, this approach incurs a significant cost to both filter area and insertion loss since the inductor's high resistive losses load the resonator's Q-factor. A similar approach can be used to enhance the magnetostatic coupling with a distinct advantage. For a MSW resonator, a series capacitor is required to compensate for $L_0$ which can be fabricated on chip with high Q-factors. Furthermore, the discrepancy between the electromagnetic and magnetostatic wavelengths is significantly smaller than that for an acoustic wave, allowing the MSW cavity's size to be on the order of the electromagnetic wavelength. As a result, an open-ended distributed element can introduce a compensating impedance with minimal cost to both area and insertion loss. This work demonstrates an order of magnitude increased effective coupling by using a distributed electromagnetic resonator as the MSW transducer with only a $68\%$ increase in resonator area at \SI{18}{\giga\hertz}. 

Fig. $\ref{chip_pictures}$ shows a chip microphotograph of the fabricated coupling-enhanced MSFVW resonators, and Fig. $\ref{chip_cartoon}$ shows a rendering of one such resonator. Each resonator consists of a rectangular $\SI{3}{\micro\meter}$-thick YIG mesa with length $\gg$ width and an open-ended $\SI{3}{\micro\meter}$-thick conformal Au transducer over the YIG. At $\SI{0}{Oe}$ bias, the transducers are designed to provide a small resistive input impedance at either $\SI{7}{\giga\hertz}$ or $\SI{18}{\giga\hertz}$. For the top row of devices with $\SI{1500}{\micro\meter}$ long Au and YIG, this is accomplished using a quarter-wave stub. The bottom row uses either $\SI{500}{\micro\meter}$ or $\SI{750}{\micro\meter}$ long YIG mesas terminated with a radial stub giving a low impedance, wide bandwidth short in a smaller form factor than the quarter-wave stub \cite{wadell_transmission_1991}. Near the electromagnetic resonance, $f_{em}$, the input impedance changes rapidly from capacitive to inductive and cannot be modeled using only the lumped inductor in Fig. $\ref{lumped_circuit}$. Instead, in Fig. $\ref{distributed_circuit}$, the inductor $L_0$ is replaced by a transmission line with impedance $Z_0$ and physical length $P_0$ (for both the quarter-wave stub and radial stub). The MSW resonance at $f_m$ is approximately modeled using a parallel R-L-C configuration \cite{asao_tunable_1995}. The interaction between the distributed resonator and the MSW will result in two anti-resonances: one below $f_{m}$ and one above \cite{psychogiou_hybrid_2015}. To account for both anti-resonances as $f_{m}$ is tuned, the coupling coefficient equation $(\ref{coupling})$ is modified as follows: $f_2=f_{m}$ if $f_m<f_{em}$ and otherwise $f_1=f_{m}$. The coupling reaches a maximum when $f_{m}=f_{em}$.

\begin{figure}[!b]
	\centering
    % \vspace*{-0.15in}
	\includegraphics[width=3.4in]{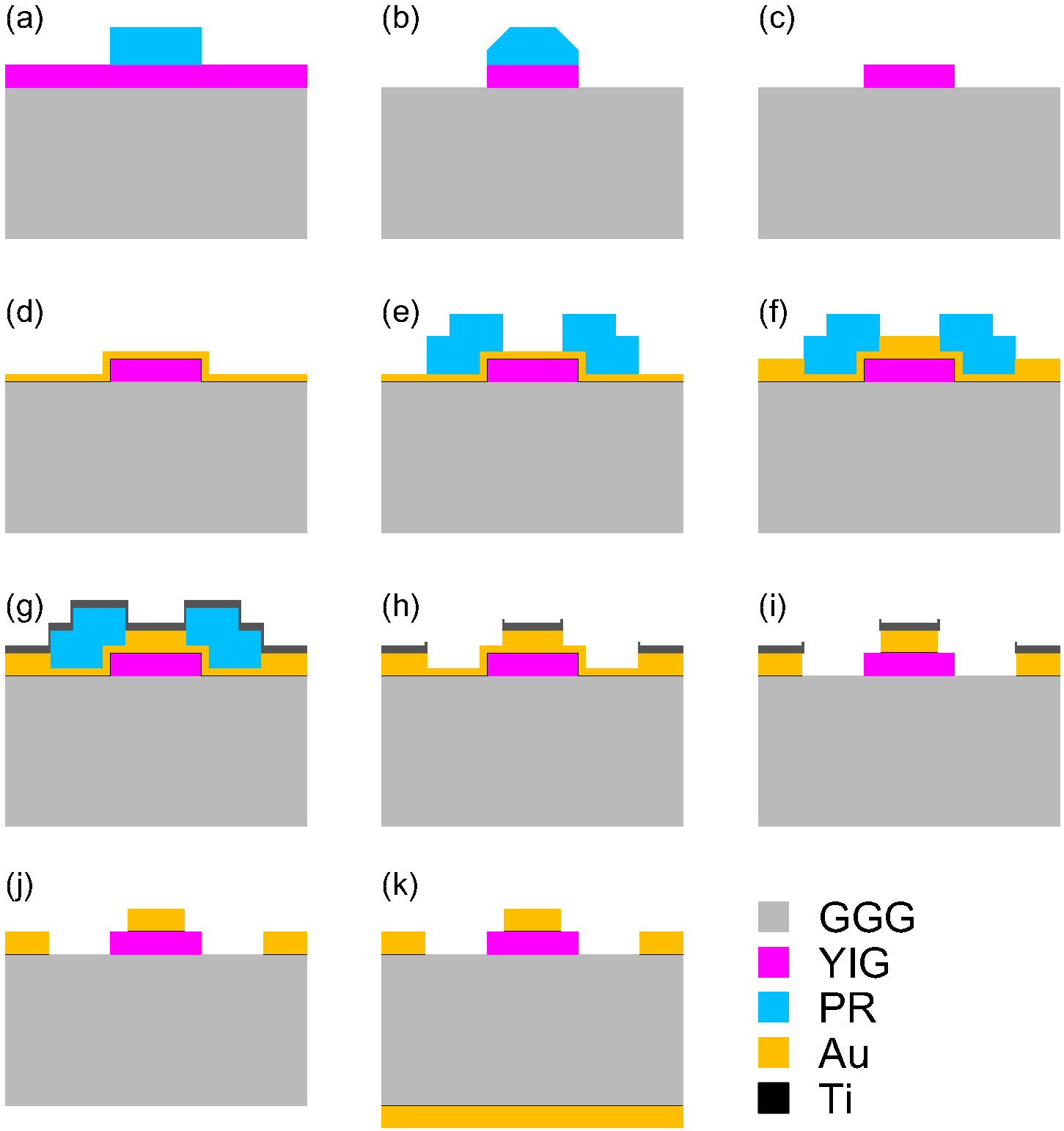}
	\caption{Fabrication process of the MSW resonator: \textbf{(a)} $\SI{7.8}{\micro\meter}$ thick photoresist (SPR220-7.0) is patterned as an etch mask on a $\SI{3}{\micro\meter}$ liquid phase epitaxy (LPE) YIG film grown on $\SI{500}{\micro\meter}$ GGG substrate. \textbf{(b)} $\SI{3}{\micro\meter}$ ion mill etch of the YIG film at a rate of $\SI{36}{\nano\meter / \minute}$. \textbf{(c)} Photoresist mask is removed, and etched YIG is soaked in phosphoric acid at $80 \, ^\circ C$ for $ \SI{20}{\min}$. \textbf{(d)} A blanket seed layer of $\SI{10}{\nano\meter}$ Ti and $\SI{300}{\nano\meter}$ Au is deposited using glancing angle e-beam evaporation. \textbf{(e)} A $\SI{10}{\micro\meter}$ thick photoresist (AZ 10XT) mask is lithographically defined. \textbf{(f)} Patterned electroplating of $\SI{2.9}{\micro\meter}$ Au. \textbf{(g)} A $\SI{150}{\nano\meter}$ Ti etch mask is deposited using glancing angle e-beam evaporation. \textbf{(h)} Liftoff of Ti mask. \textbf{(i)} Au etch to remove blanket seed layer. \textbf{(j)} Ti etch. \textbf{(k)} Blanket $\SI{10}{\nano\meter}$ Ti and $\SI{300}{\nano\meter}$ Au is evaporated on the bottom of the substrate followed by electroplating of $\SI{3.0}{\micro\meter}$ of Au.}
    \label{fig_fab_process}
     % \vspace*{-0.15in}
\end{figure}

\section{Fabrication}
\label{sec_fab}

\begin{figure}[!t]
	\centering
 
	% \vspace*{-0.15in}
	\subfloat[]{\includegraphics[width=3.4in]{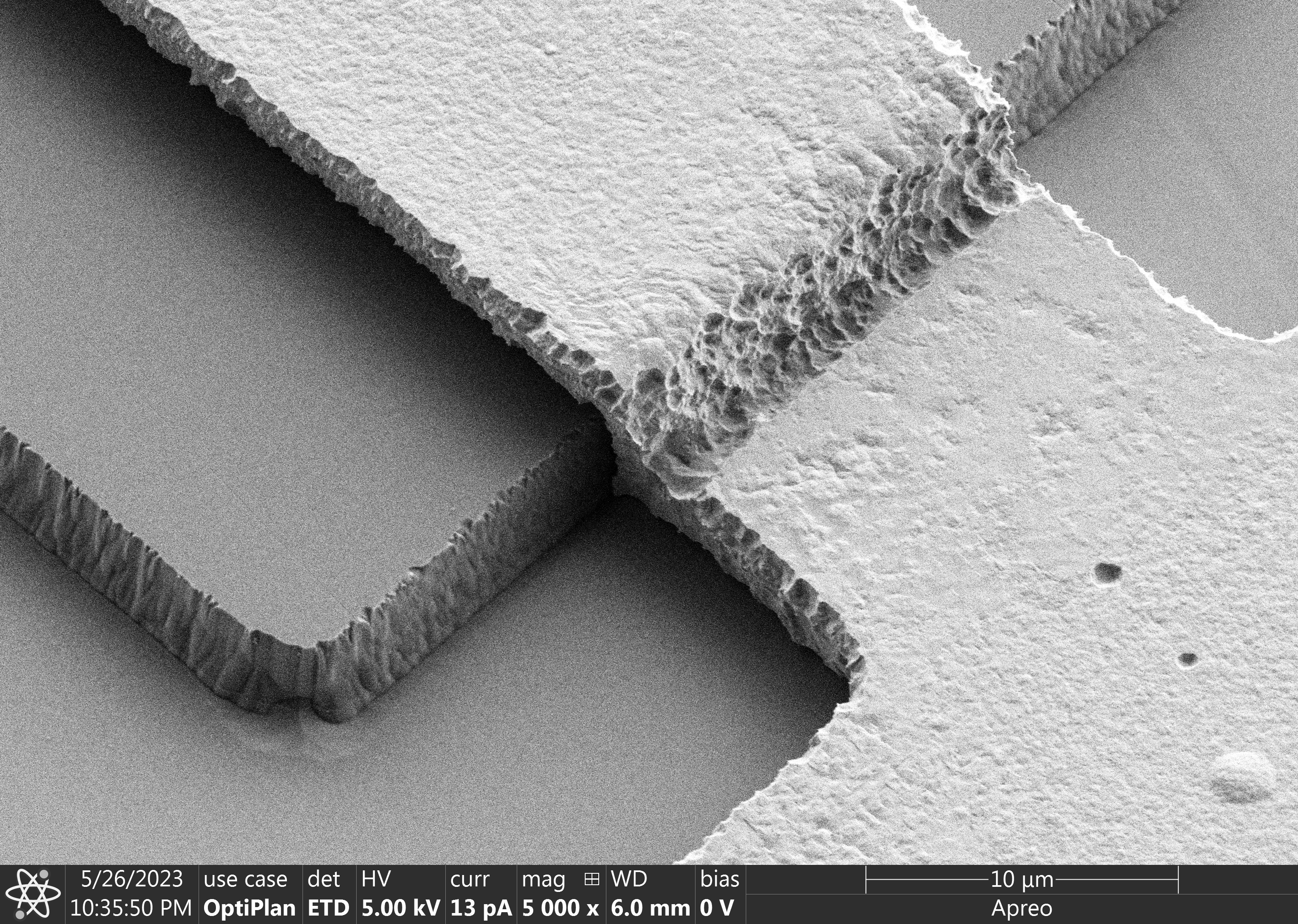}%
		\label{sem_electrode_initial_mask}}
%	\hfil

	\subfloat[]{\includegraphics[width=3.4in]{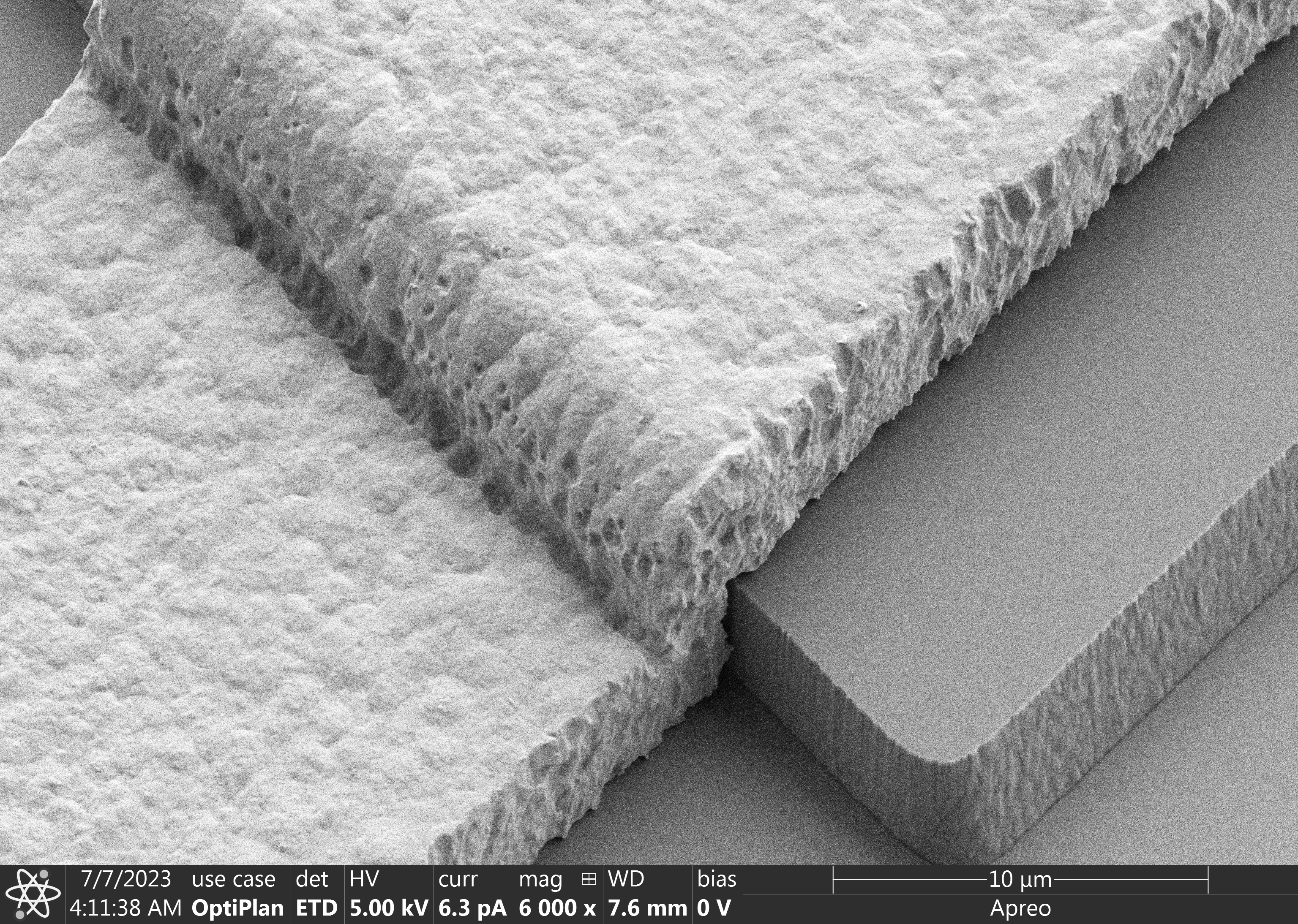}%
		\label{sem_electrode_better_mask}}
	
	\caption{Etched YIG and electroplated Au electrodes after final Au and Ti wet etches using \textbf{(a)} $\SI{70}{\nano\meter}$ e-beam evaporated Ti mask (not glancing angle) and \textbf{(b)} thicker $\SI{150}{\nano\meter}$ glancing angle e-beam evaporated Ti mask. In both cases, the buffered oxide etchant undercuts the Ti adhesion layer beneath the electroplated Au.}
	% \vspace*{-0.15in}
	\label{SEM_transducer_step_coverage}
\end{figure}

The fabrication process for the MSFVW resonators is outlined in Fig. $\ref{fig_fab_process}$. The YIG etching in steps (a)-(c) is adapted from and described in detail in \cite{feng_micromachined_2023, dai_octave-tunable_2020, devitt}. Starting with a $\SI{3}{\micro\meter}$ YIG film epitaxially grown on a $\SI{500}{\micro\meter}$ thick GGG substrate, the YIG mesas are patterned through ion milling using a thick photoresist (PR) mask. After ion milling, the sample is cleaned and soaked in hot phosphoric acid to remove the re-sputtered material. At this point, the MSW resonator is defined, but the microwave transducers must still be deposited. Based on \cite{devitt}, $\SI{300}{\nano\meter}$ thick Au transducers showed significant conductor losses so the process presented here is optimized for thick Au transducers. A patterned electroplating process \cite{zhang_process_2004} is developed as a cost effective alternative to evaporation and lift-off \cite{pulskamp_monolithically_2009}. A $\SI{10}{\nano\meter}$ Ti adhesion layer and blanket $\SI{300}{\nano\meter}$ Au seed layer is first evaporated at a glancing angle to ensure coverage on the sidewalls of the etched YIG. The transducer geometry is lithographically defined using $\SI{10}{\micro\meter}$ thick AZ-10XT photoresist chosen for compatibility with the electroplating solution. Using the blanket seed layer as an electrode, $\SI{2.9}{\micro\meter}$ Au is electroplated around the photoresist pattern. To protect the thick Au transducers during the seed layer removal, a Ti wet etch mask is evaporated and patterned by liftoff of the photoresist mask. The sample is submerged in Au enchant to remove the seed layer followed by buffered oxide etchant (BOE) to remove the Ti adhesion layer and etch mask. Initially, the Ti mask was $\SI{70}{\nano\meter}$ thick and evaporated at normal incidence yielding the transducer in Fig. $\ref{sem_electrode_initial_mask}$ after the Au/Ti wet etches. The top surface of the plated Au shows randomly distributed pitting indicating pin holes in the mask and the etchant attacked the Au on the edge of the YIG mesa since the evaporated Ti mask was not conformal. In the next device iteration, a thicker $\SI{150}{\nano\meter}$ Ti mask was used to minimize any pin holes and glancing angle e-beam evaporation was used to ensure conformal Ti coverage over the edge of the YIG mesa. Additionally, the Ti adhesion layer beneath the electroplated Au is undercut by the BOE. Lastly, $\SI{10}{\nano\meter}$ Ti and $\SI{300}{\nano\meter}$ Au is evaporated followed by $\SI{3}{\micro\meter}$ Au electroplating on the back-side of the chip.

\section{Resonator Measurement}

\begin{figure}[!t]
	\centering
	
	% \vspace*{-0.05in}
	\subfloat[]{\includegraphics[width=2.8in]{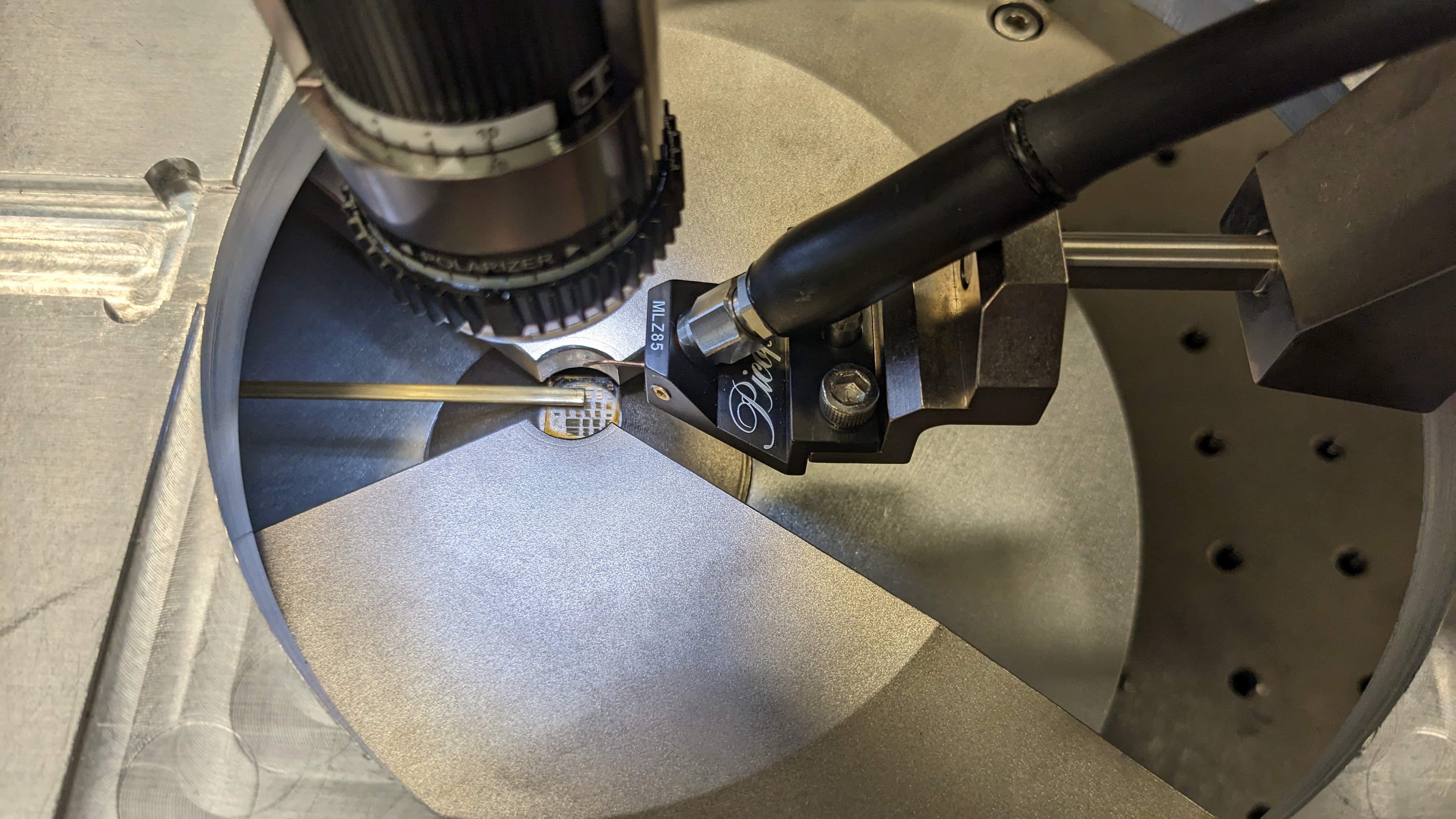}%
		\label{measurement_setup_1}}
	% \hfil
 
	\subfloat[]{\includegraphics[width=2.8in]{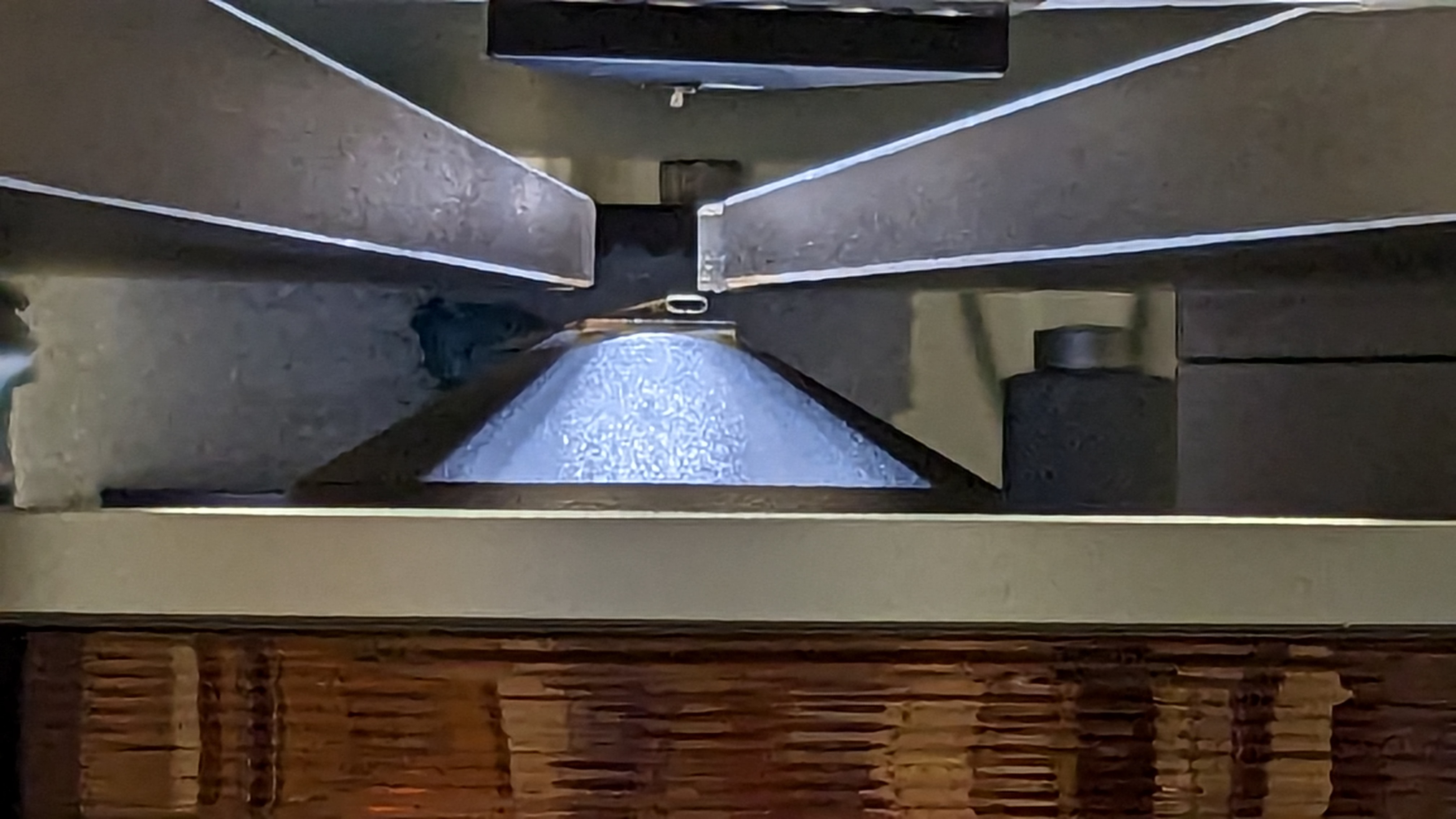}%
		\label{measurement_setup_2}}
	
        \caption{Photos of the measurement setup for the fabricated $\SI{10}{\milli\meter}\times\SI{10}{\milli\meter}$ resonator chip. \textbf{(a)} Top down view of the setup showing the optical microscope, a non-magnetic GSG probe, and a Gauss meter in close proximity to the chip resting on the electromagnet. \textbf{(b)} Side view of the setup with the chip resting flat on the pole of the electromagnet.}

        % \vspace*{-0.15in}
 
	\label{measurement_setup}
\end{figure}

\begin{figure}[!t]
	\centering
	%	\captionsetup[subfigure]{labelformat=empty}
	
	\vspace*{-0.1in}
	% \hspace*{-0.15in}s
	\subfloat[]{\includegraphics[width=3.4in, valign=c]{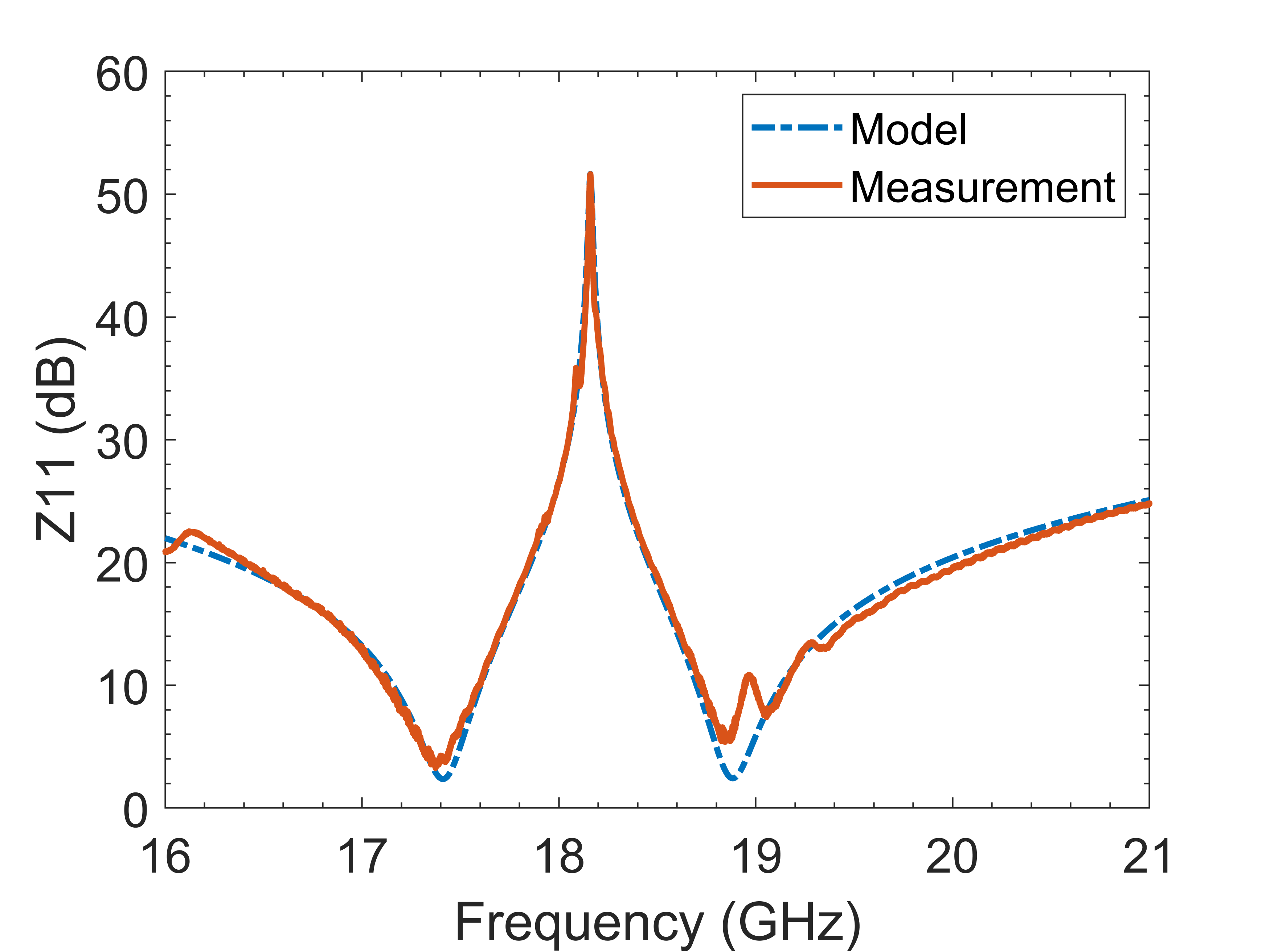}%
		\label{Model_fit_z11}}
	% \hspace*{-0.15in}
        % \vspace*{-0.15in}
        
	\subfloat[]{\includegraphics[width=3.4in, valign=c]{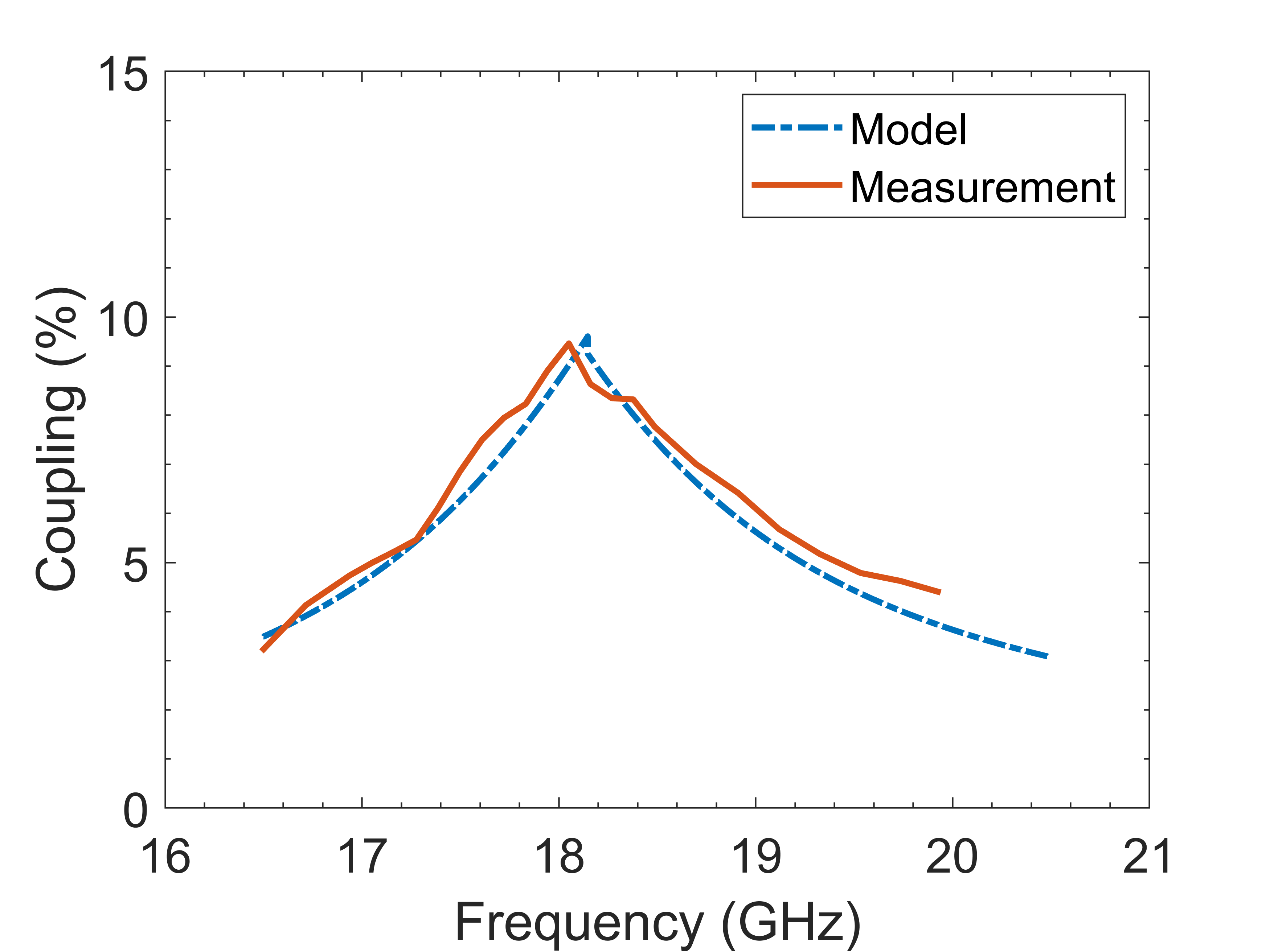}%
		\label{Model_fit_coupling}}
	\\

	\subfloat[]{\adjustbox{valign=c, width=3.4in}{\begin{tabular}{|c|c|c|c|c|c|c|}\hline
				\multicolumn{7}{|c|}{\textbf{Model Parameters}} \\ \hline
				$R_0$ & $Z_0$ & $P_0$ & $R_m$ & $L_m$ & $f_m$ & $\varepsilon_r$\\ \hline 
				$\SI{1.256}{\ohm}$ &  $\SI{75.51}{\ohm}$ & $\SI{1622}{\micro\meter}$ & $\SI{381.7}{\ohm}$ & $\SI{3.414}{\pico\henry}$ & $\SI{18.16}{\giga\hertz}$ & 6.5 \\ \hline
		\end{tabular}}%
		\label{Model_params_table}}

	\caption{\textbf{(a)} Measured resonator impedance response biased at $\SI{7714}{Oe}$  showing a Q-factor of 1274 and simulated response from the fitted distributed circuit model. \textbf{(b)} Measured resonator coupling from $\SI{7163}{Oe}$ to $\SI{8310}{Oe}$ and simulated model coupling sweeping $f_m$ from $\SI{16.5}{\giga\hertz}$ to $\SI{20.5}{\giga\hertz}$. \textbf{(c)} Fitted distributed circuit model parameters at a $\SI{7714}{Oe}$ bias.}
	
	% \vspace*{-0.15in}
	\label{Model_fitting}
\end{figure}

1-port s-parameters are measured using an Agilent PNA E8364B with non-magnetic GSG probes at an input power of $\SI{-15}{\decibel m}$. The out-of-plane magnetic field is generated using a constant current source and a single-pole electromagnet with a $\SI{10}{\milli\meter}$ pole face diameter. A single-axis Gauss meter attached to a probe manipulator is positioned closely above the sample near the device under test (DUT) to estimate the magnitude of the applied field. To ensure a uniform bias field across the entire length of the YIG mesa, the DUT is aligned to the center of the pole face using an optical microscope and secured in place using an adhesive. Fig. $\ref{measurement_setup}$ illustrates the measurement setup for the MSW resonators.

\begin{figure}[!t]
	\centering
	
	\vspace*{-0.15in}
	\subfloat[]{\includegraphics[width=3.4in]{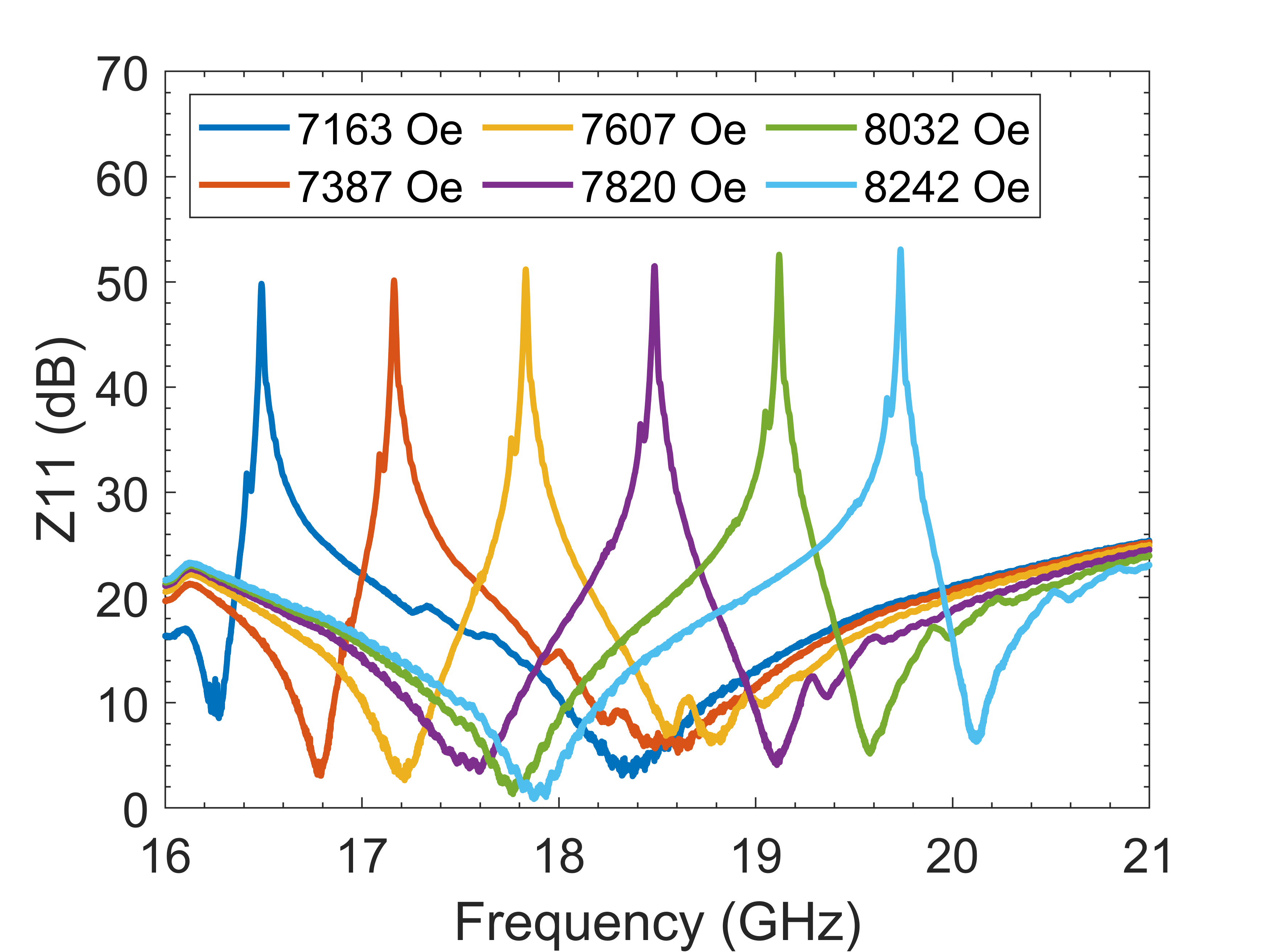}%
		\label{18ghz_resonator_z11}}
	% \hspace*{-0.17in}
        \vspace*{-0.15in}

        \subfloat[]{\includegraphics[width=3.4in]{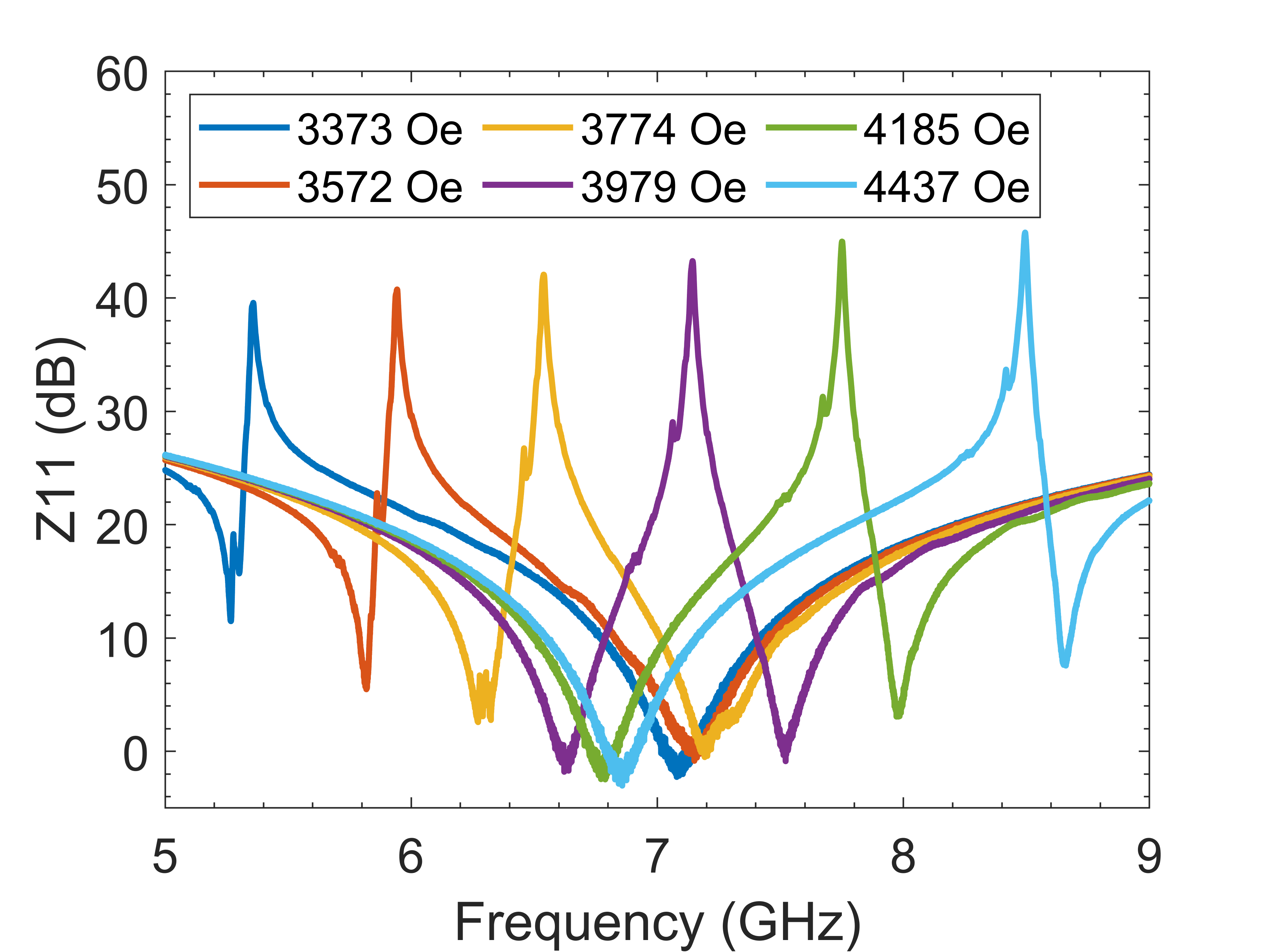}%
		\label{7ghz_resonator_z11}}

        \caption{ Measured impedance over bias field for the \textbf{(a)} $\SI{18}{\giga\hertz}$ and \textbf{(b)} $\SI{7}{\giga\hertz}$ resonators.}

	\label{resonator_Z11}
	\vspace*{-0.15in}
\end{figure}

    \begin{figure}[!b]
        \centering
        
        \vspace*{-0.2in}
        \includegraphics[width=3.4in]{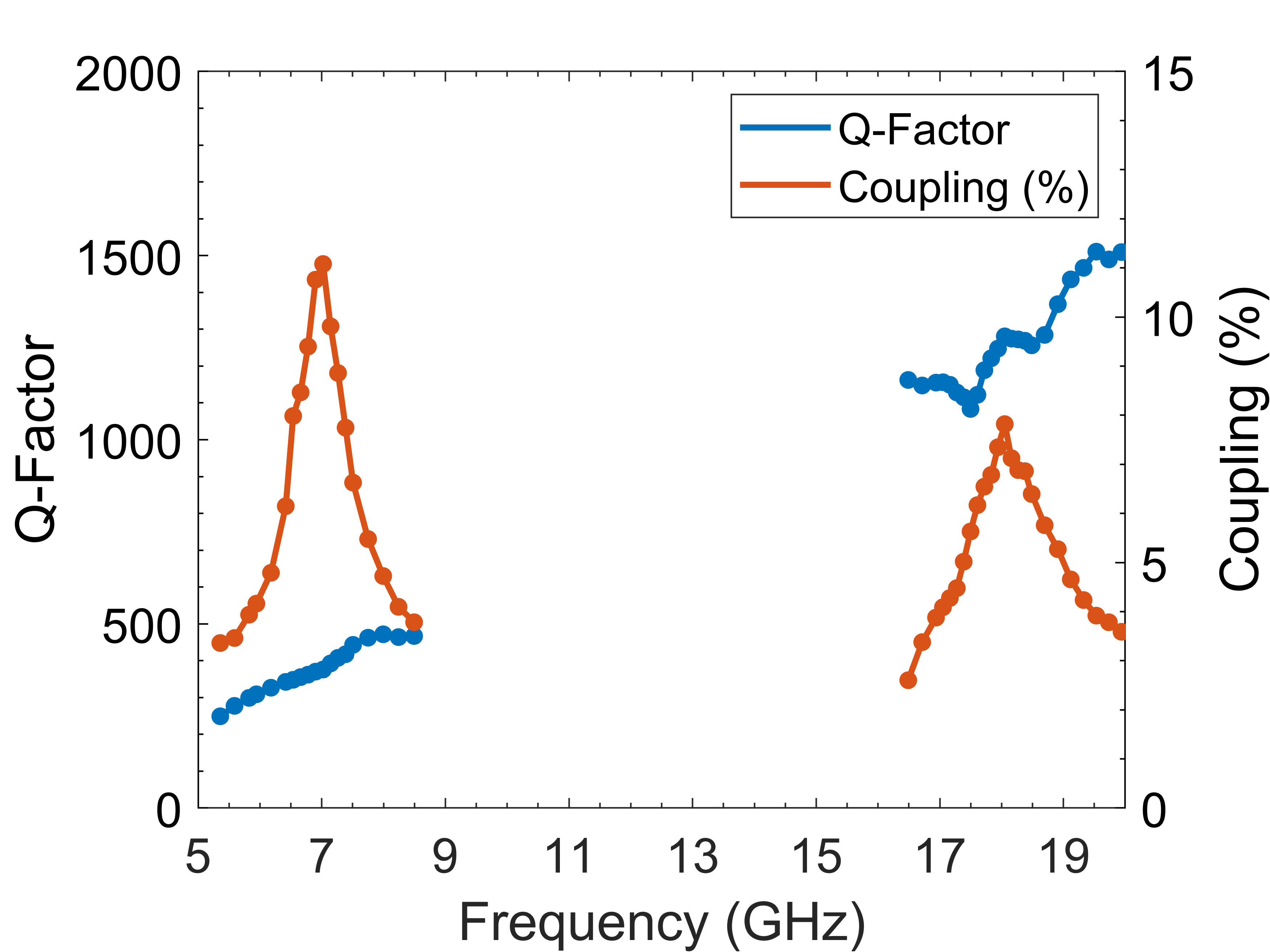}%
        
        \caption{Extracted Q-factor and coupling over bias field for the $\SI{7}{\giga\hertz}$ and $\SI{18}{\giga\hertz}$ resonators.}
        
        \label{resonator_coupling_Q}
        \vspace*{-0.1in}
    \end{figure}

\begin{figure}[!t]
	\centering
	
	\vspace*{-0.15in}
	% \hspace*{-0.10in}
	\subfloat[]{\includegraphics[width=3.4in]{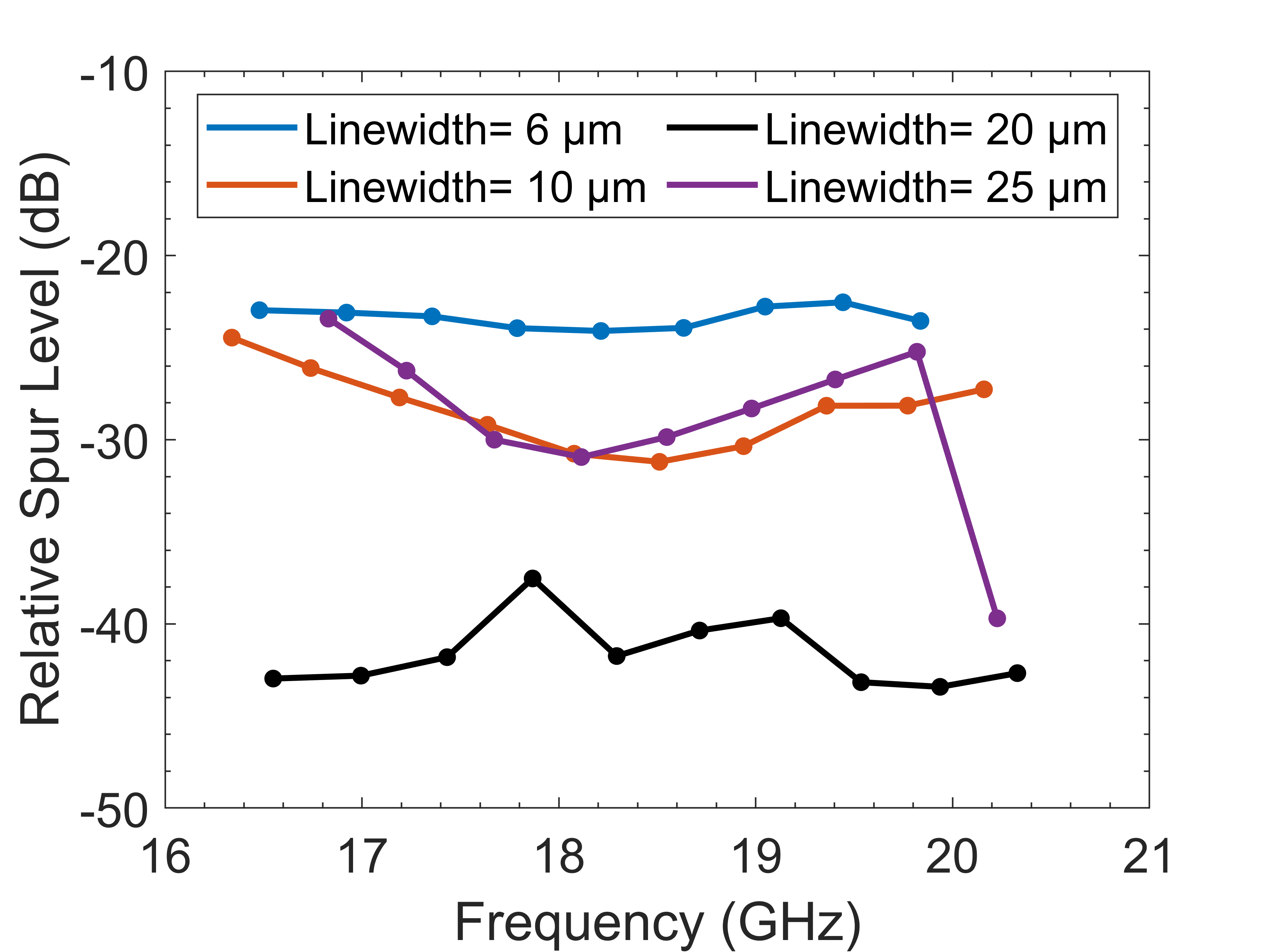}%
		\label{linewidth_spurs}}
	% \hspace*{-0.17in}
        \vspace*{-0.15in}
        
	\subfloat[]{\includegraphics[width=3.4in]{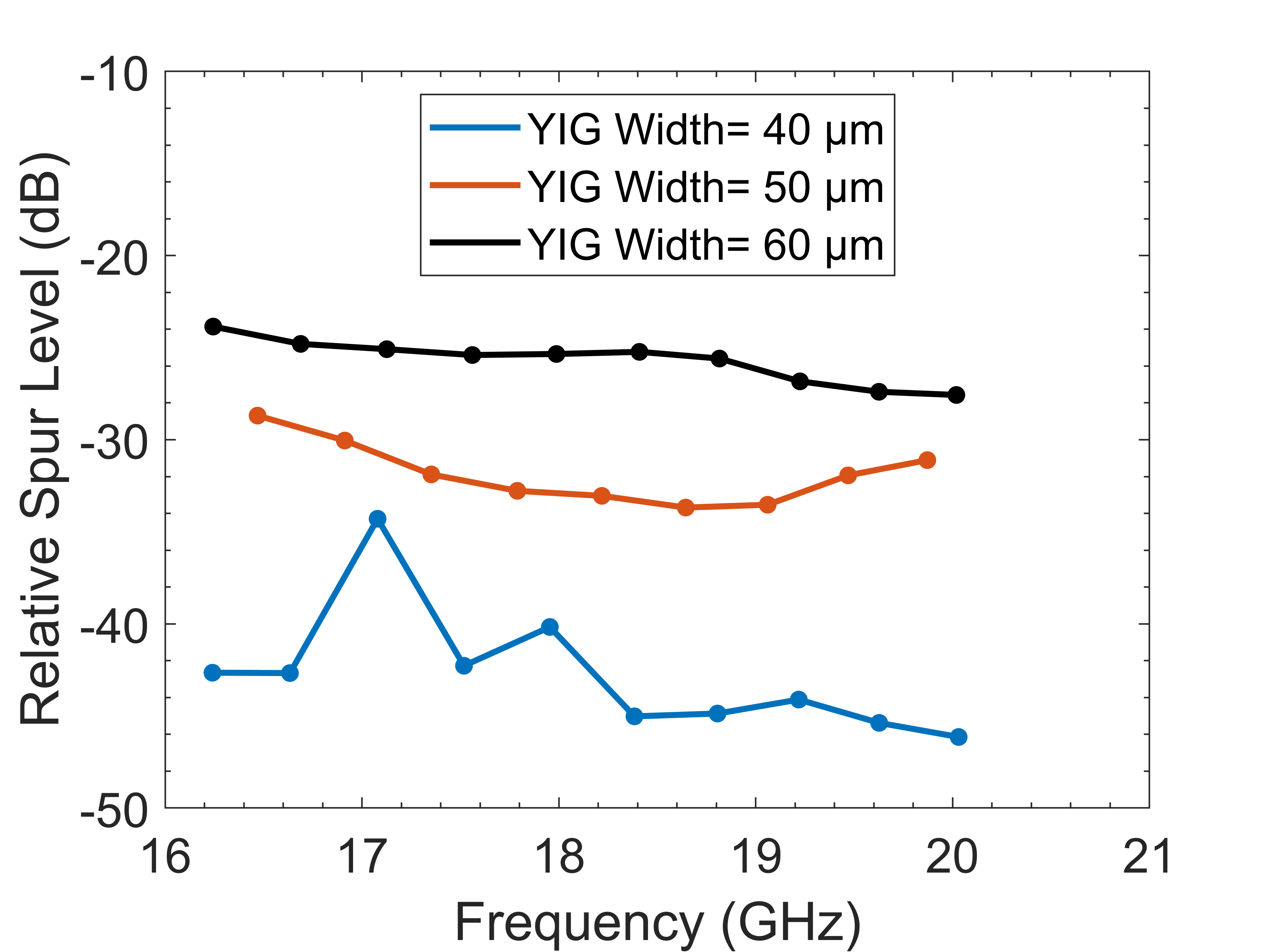}%
		\label{yig_width_spurs}}
	
	\caption{Measured impedance of width-mode spur level relative to the main resonance over resonant frequency. \textbf{(a)} YIG resonator width is fixed to $\SI{40}{\micro\meter}$ and the line width of the gold transducer is varied. \textbf{(b)} Gold transducer width is fixed to $\SI{15}{\micro\meter}$ and the YIG resonator width is varied.}
	
	\label{spur_levels}
        \vspace*{-0.15in}
\end{figure}
\begin{figure}[!t]
	\centering
	
	\vspace*{-0.15in}
	% \hspace*{-0.10in}
	\subfloat[]{\includegraphics[width=3.4in]{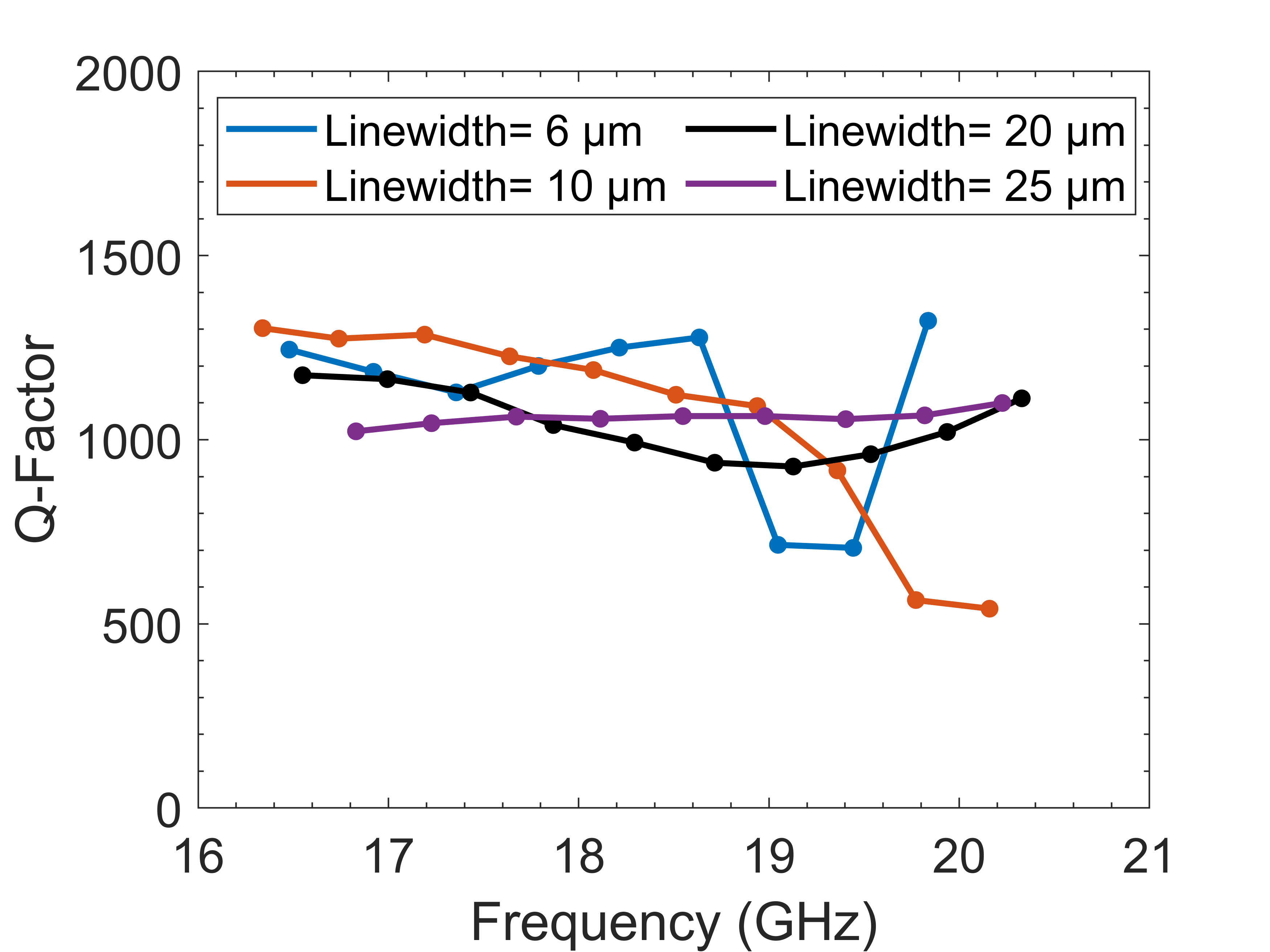}%
		\label{linewidth_q}}
	% \hspace*{-0.17in}

        % \vspace*{-0.15in}
        
	\subfloat[]{\includegraphics[width=3.4in]{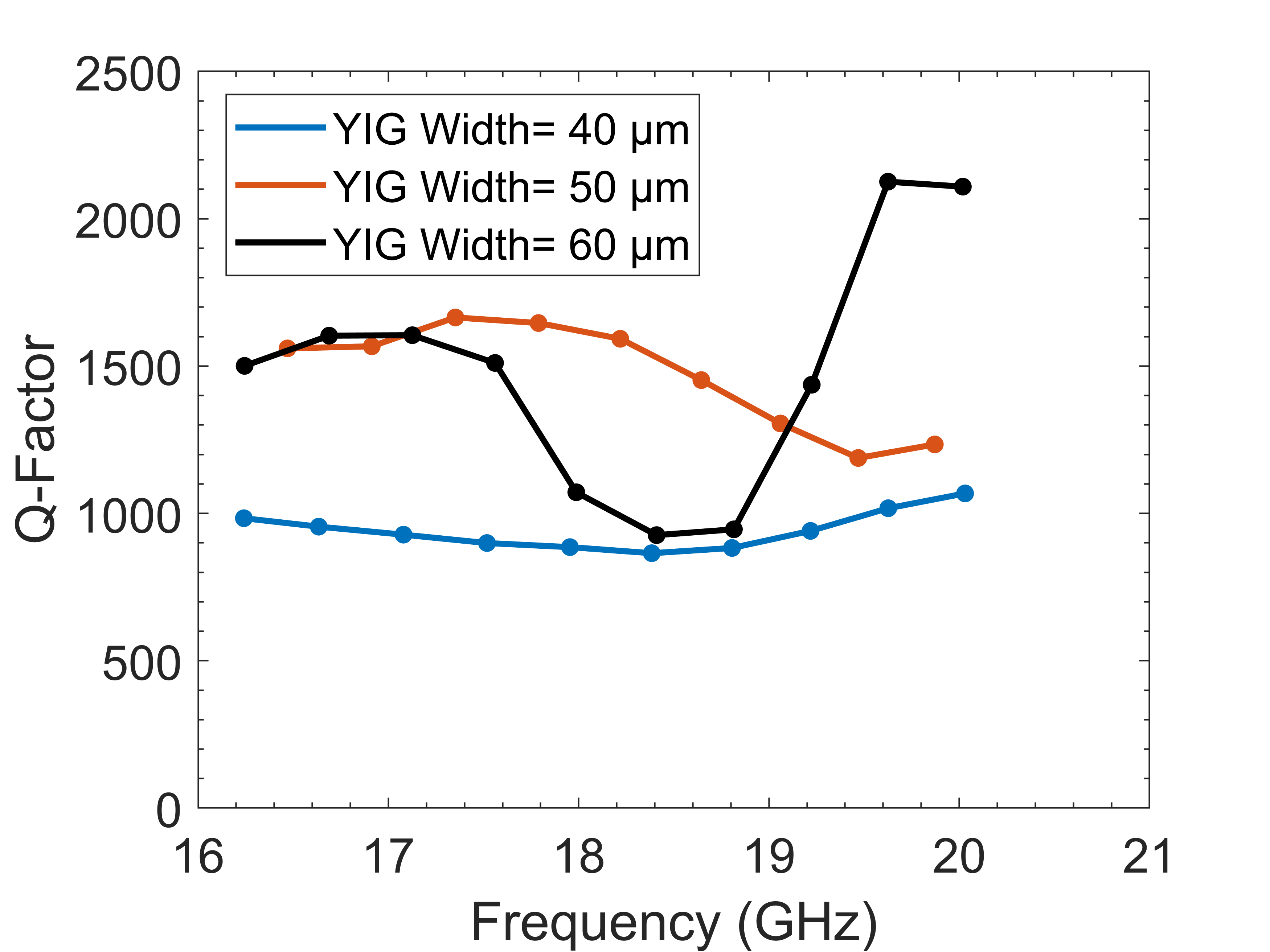}%
		\label{yig_width_q}}
	
	\caption{Measured Q-factor over resonant frequency. \textbf{(a)} YIG resonator width is fixed to $\SI{40}{\micro\meter}$ and the line width of the gold transducer is varied. \textbf{(b)} Gold transducer width is fixed to $\SI{15}{\micro\meter}$ and the YIG resonator width is varied.}
	
	\label{qFactor_trends}
        % \vspace*{-0.15in}
\end{figure}

\begin{figure}[!b]
	\centering
	\begin{circuitikz}[scale=0.65]
		\ctikzset{label/align=rotate}
		
		\ctikzset{capacitors/scale=0.4}
		\ctikzset{resistors/scale=0.5}
		\ctikzset{inductors/scale=0.65}
		
		\draw[<->,>=stealth] (0,-2.25) to [bend left] node[pos=0.5,fill=white] {$M_{12}$} ++(4,0);
		\draw[<->,>=stealth] (4,-2.25) to [bend left] node[pos=0.5,fill=white] {$M_{23}$} ++(4,0);

            \draw [black,domain=-135:-45] plot ({0.5*cos(\x)+3.5}, {0.5*sin(\x) -7.5 -0.5*sin(-135)});
            \draw[black](3.5,-7) -- ({3.5-0.25*sqrt(2)}, -7.5);
            \draw[black](3.5,-7) -- ({3.5+0.25*sqrt(2)},-7.5);

            \draw [black,domain=-135:-45] plot ({0.5*cos(\x)+4.5}, {0.5*sin(\x) -7.5 -0.5*sin(-135)});
            \draw[black](4.5,-7) -- ({4.5-0.25*sqrt(2)}, -7.5);
            \draw[black](4.5,-7) -- ({4.5+0.25*sqrt(2)},-7.5);
            
        \node at (4,-4.75) {$M_{IO}$};
  
		\draw
		
		(0,0) to[generic= $Z_{qwt}$] (0,2)
		(8,0) to[generic= $Z_{qwt}$] (8,2)
		
		(0,-2) to[R=$R_0$] (0,0)
		(8,-2) to[R=$R_0$] (8,0)
		
		(-1,-4) to[R=$R_{m1}$] (-1,-2)
		(0,-4) to[L=$L_{m1}$] (0,-2)
		(1,-4) to[C=$C_{m1}$] (1,-2)

		(3,-4) to[R=$R_{m2}$] (3,-2)
		(4,-4) to[L=$L_{m2}$] (4,-2)
		(5,-4) to[C=$C_{m2}$] (5,-2)

		(7,-4) to[R=$R_{m3}$] (7,-2)
		(8,-4) to[L=$L_{m3}$] (8,-2)
		(9,-4) to[C=$C_{m3}$] (9,-2)

		(3.5,-7) to[generic= $Z_0${$,\,$}$P_0$] (3.5,-5)
		(4.5,-5) to[generic= $Z_0${$,\,$}$P_0$] (4.5,-7)
		
		% (3.5,-6) to[open, -o] (3.5,-7)
		% (4.5,-6) to[open, -o] (4.5,-7)

		(-1,-2) -- (1,-2)
		(-1,-4) -- (1,-4)
		(3,-2) -- (5,-2)
		(3,-4) -- (5,-4)
		(7,-2) -- (9,-2)
		(7,-4) -- (9,-4)
		
		(0,-4) -- (0,-5)
		(0,-5) -- (3.5,-5)
		
		(8,-4) -- (8,-5)
		(8,-5) -- (4.5,-5)
		
		(3.75,-5.5) -- (4.25,-6.5)
		(3.75,-6.5) -- (4.25,-5.5)
		;
	\end{circuitikz}
	\caption{3\textsuperscript{rd}-order MSW filter topology leveraging edge-coupled YIG resonators and quarter-wave impedance matching sections at the input and output.}
	\label{filter_topology}
    % \vspace*{-0.2in}
\end{figure}
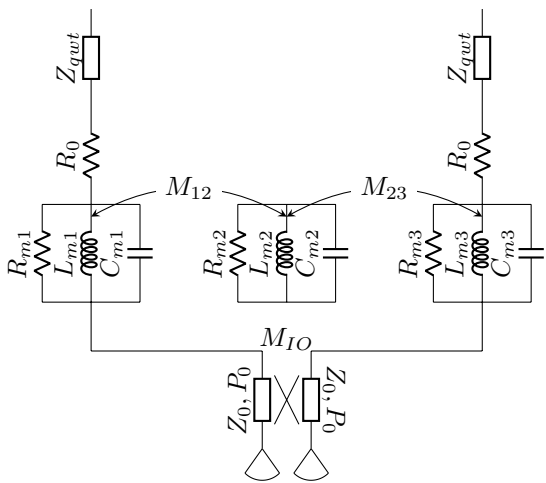

Fig. $\ref{Model_fit_z11}$ shows the frequency response of the \SI{18}{\giga\hertz} resonator (highlighted in Fig. \ref{chip_pictures}) terminated with a radial stub biased at $\SI{7714}{Oe}$ along with the simulated distributed circuit model using the fitted parameters listed in Fig. $\ref{Model_params_table}$. The distributed model shows good agreement with the measured frequency response and accurately captures the frequency dependence of the coupling as shown in Fig. $\ref{Model_fit_coupling}$ by sweeping $f_m$. A Smith plot of the resonator response is provided in the appendix. Figures $\ref{resonator_Z11}$ and $\ref{resonator_coupling_Q}$ show the measured frequency response, quality factors, and coupling factors of the two characteristic devices highlighted in Fig. \ref{chip_pictures} designed at \SI{18}{\giga\hertz} and \SI{7}{\giga\hertz}. In both, the YIG mesa is $\SI{40}{\micro\meter}$ wide and $\SI{750}{\micro\meter}$ long while the Au transducer is $\SI{20}{\micro\meter}$ wide. Only the designed frequency of the radial stub differs between the two. Finite element simulations revealed that increasing the width of the Au transducers relative to the YIG width helped suppress higher order MSFVW width modes, so two sets of resonators were fabricated. The first with YIG dimensions of $\SI{40}{\micro\meter}$ by $\SI{1500}{\micro\meter}$ and gradually increasing transducer widths from $\SI{6}{\micro\meter}$ to $\SI{25}{\micro\meter}$ and the second with a constant $\SI{15}{\micro\meter}$ transducer width and a YIG length of $\SI{1500}{\micro\meter}$ with gradually increasing YIG widths from $\SI{40}{\micro\meter}$ to $\SI{60}{\micro\meter}$. The resulting relative spur levels defined as the difference in Z11 (in dB) between the main resonance and the first MSFVW width mode are summarized in Fig. $\ref{spur_levels}$. Based on these results optimal spur suppression is achieved when the ratio of transducer width to YIG width is between $0.375$ and $0.5$. Fig. $\ref{qFactor_trends}$ summarizes the trends in the Q-factor over YIG width and transducer width. Fig. $\ref{linewidth_q}$ shows no clear trend in Q-factor over transducer width. Fig. $\ref{yig_width_q}$ indicates that a wider YIG mesa can substantially improve Q-factor. The resonator coupling factor exhibited an increasing trend as the YIG mesa width decreased. The coupling also shows an increasing trend for wider transducer widths up to $\SI{20}{\micro\meter}$ beyond which it decreases. Several resonators on chip including the device with YIG width of $\SI{60}{\micro\meter}$ show a peak splitting and broadening above $\SI{18}{\giga\hertz}$ leading to degraded Q-factors.

\section{Filter Design and Simulation}

\begin{figure}[!t]
	\centering
	
	\vspace*{-0.15in}
	\includegraphics[width=3.4in]{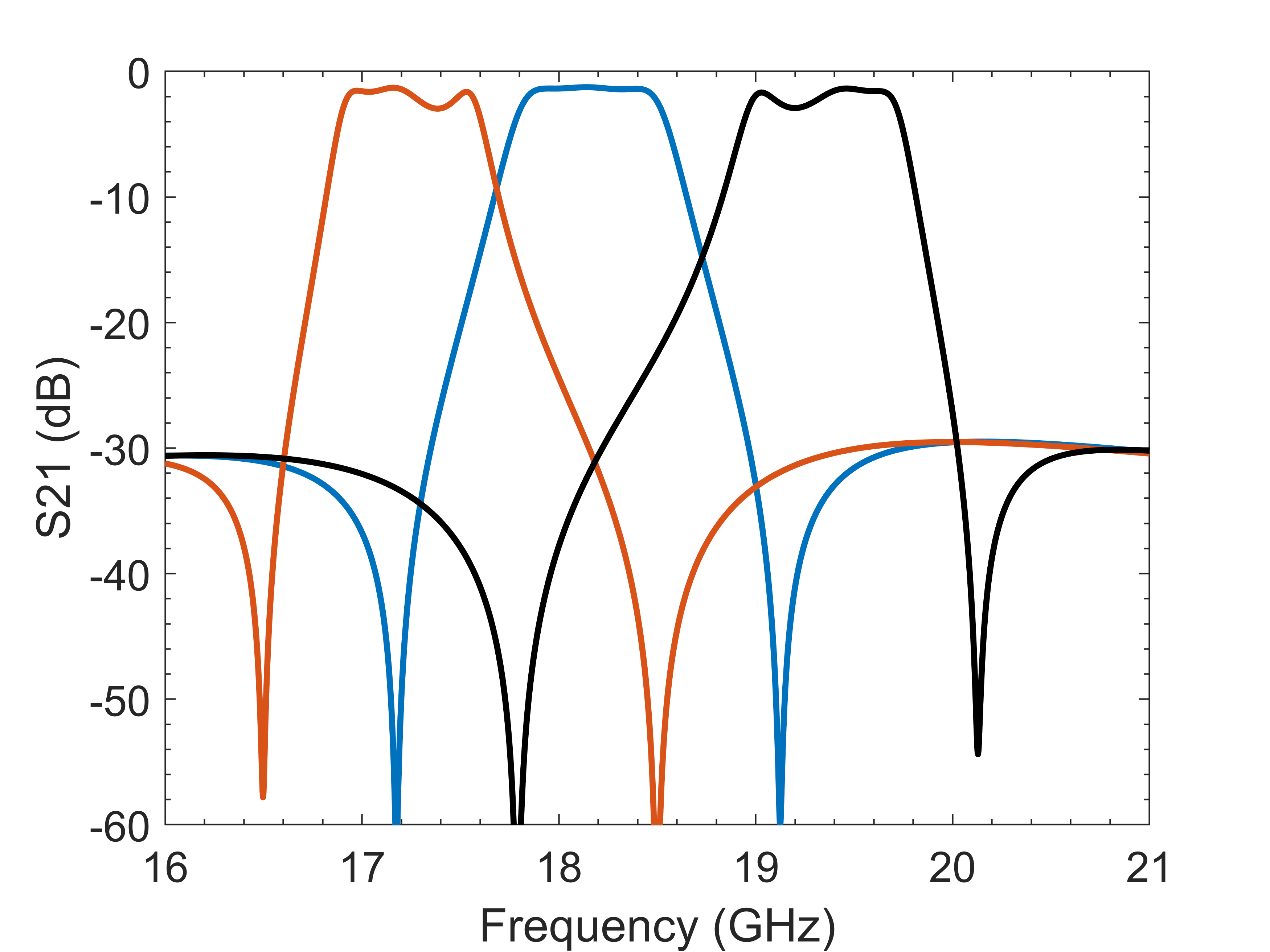}%
	
	\caption{Simulated filter responses at three different biases for the topology shown in Fig. $\ref{filter_topology}$ using the fitted model parameters listed in Fig. $\ref{Model_params_table}$. The filter is optimized for $\SI{2}{\giga\hertz}$ center frequency tuning and shows insertion loss less than $\SI{3}{\decibel}$ over a $\SI{3.98}{\%}$ fractional bandwidth.}
	
	\label{simulated_tune_filter}
        \vspace*{-0.05in}
\end{figure}

\begin{figure}[!b]
	\centering
	
	\vspace*{-0.15in}
	\subfloat[]{\includegraphics[width=3.4in]{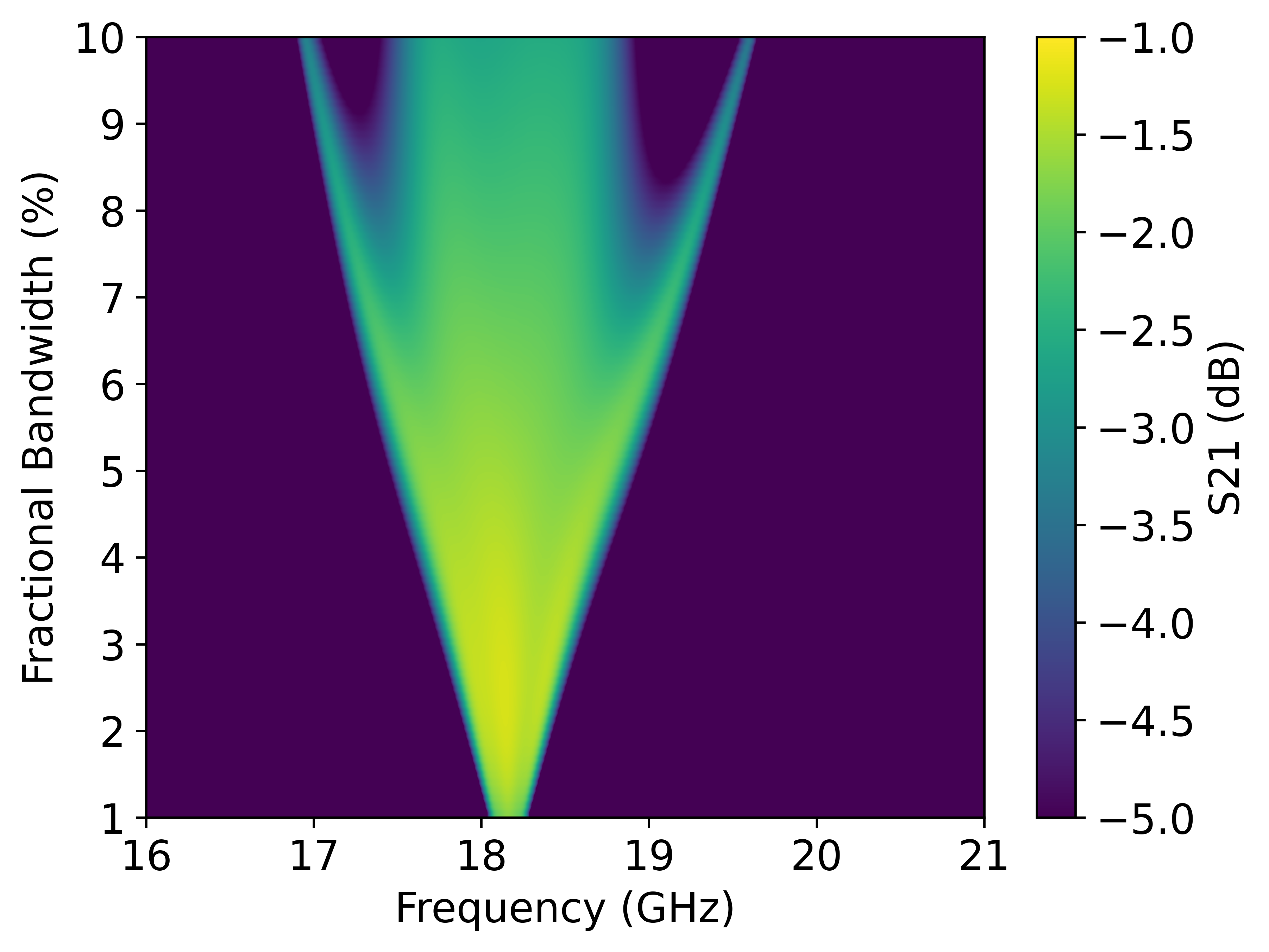}%
		\label{simulated_filter_fbw_sweep}}
	%	\hfil

    \vspace*{-0.1in}
	\subfloat[]{\includegraphics[width=3.4in]{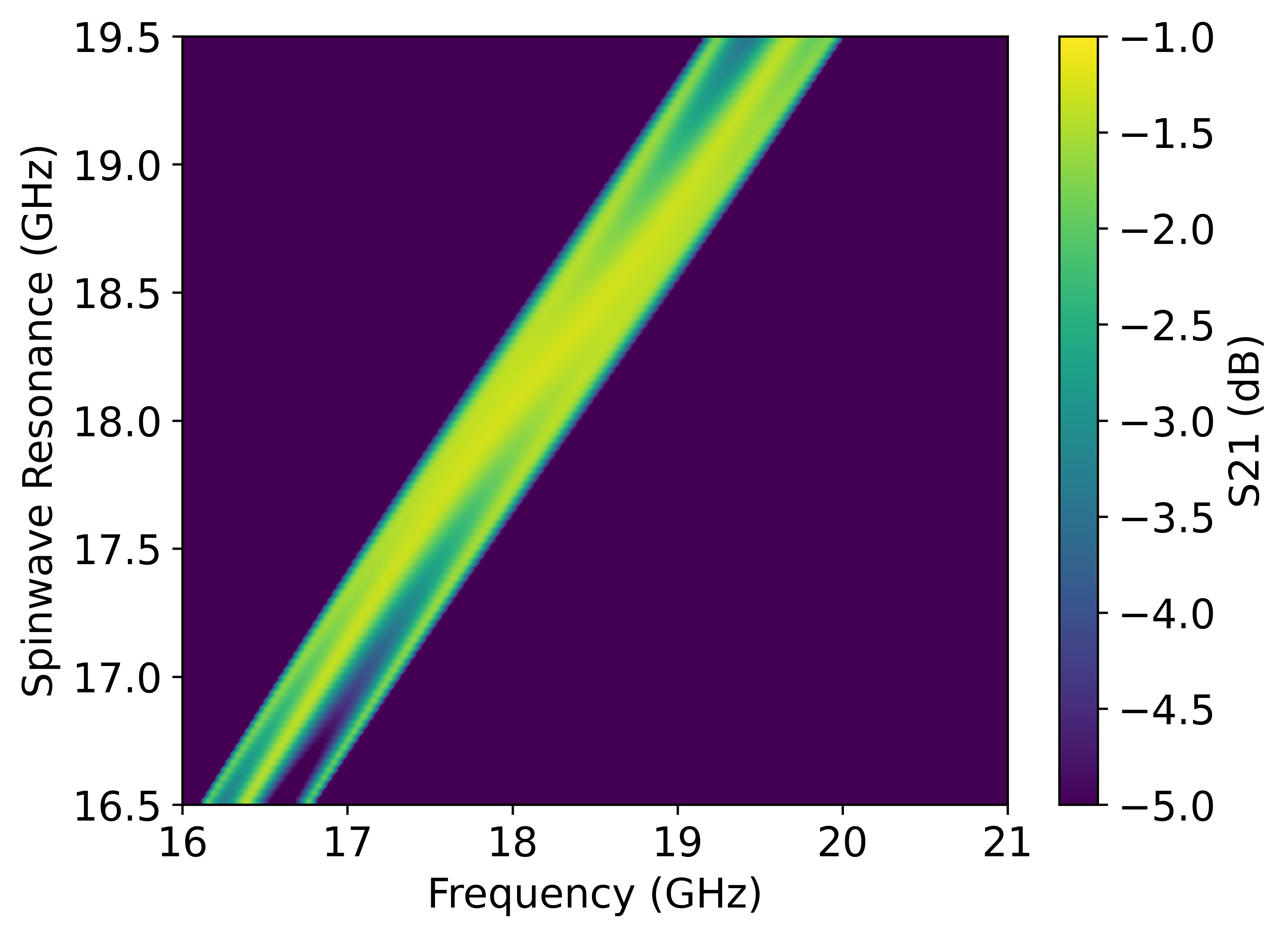}%
		\label{simulated_filter_tune_sweep}}
	
	\caption{Simulated filter insertion loss of the topology shown in Fig. $\ref{filter_topology}$ using the fitted model parameters listed in Fig. $\ref{Model_params_table}$. \textbf{(a)} The synthesized fractional bandwidth is swept while the spin wave resonance is fixed around the maximum coupling $f_m=f_{em}$. \textbf{(b)} The spin wave resonance, $f_m$, is swept while maintaining a $3\%$ synthesized fractional bandwidth.}
	
	\label{simulated_filter_sweeps}
\end{figure}

Based on the results of \cite{devitt}, MSW resonators in close proximity can couple through the magnetic field enabling the synthesis of bandpass filters using optimized inter-resonator coupling factors $M_{ij}$ analogous to the design of coupled cavity filters \cite{snyder_emerging_2021, pozar_microwave_2012, swanson_narrow-band_2007}. Fig. $\ref{filter_topology}$ demonstrates an example $3\textsuperscript{rd}$-order edge-coupled bandpass filter topology using the novel coupling-enhanced distributed MSW resonators. The filter consists of three YIG resonators with inter-resonator couplings $M_{12}$ and $M_{23}$. The first and third resonators each have a transducer to excite a MSW with a frequency-dependent coupling coefficient $k_{eff}^2(f_m)$. The transducers are modeled using coupled transmission lines to account for their finite electromagnetic coupling, $M_{IO}$. Two quarter-wave sections with impedance $Z_{qwt}$ are used to match the $\SI{50}{\ohm}$ port to the resonators. Fig. $\ref{simulated_tune_filter}$ shows a sample frequency response using the model parameters listed in $\ref{Model_params_table}$. The $\SI{3.98}{\%}$ fractional bandwidth filter is optimized for center frequency tuning over a $\SI{2}{\giga\hertz}$ range while maintaining less than $\SI{3}{\decibel}$ insertion loss.  Since $k_{eff}^2(f_m)$ varies over frequency with a maximum at $f_m=f_{em}$, the filter design must make a trade-off between maximum bandwidth and center frequency tuning range. Both the insertion loss and ripple increase as either the center frequency ($f_m$) is tuned away from $f_{em}$ or the synthesized bandwidth is increased. Fig. $\ref{simulated_filter_sweeps}$ illustrates this design space showing the simulated insertion loss of the filter while sweeping both the synthesized bandwidth and magnetic tuning. The resonator parameters listed in Fig. $\ref{Model_params_table}$ are held constant.

The finite bandwidth of the quarter-wave sections will also limit the maximum tuning and bandwidth, but the magnetostatic coupling coefficient dominates performance for these resonators. The required inter-resonator coupling is proportional to the fractional bandwidth \cite{dishal_alignment_1951, swanson_narrow-band_2007} and is allowed to be arbitrarily large ($M_{ij}>3\%$) in the filter simulations. Achieving such large couplings for wider bandwidths represents a micromachining challenge.

\section{Conclusion}

\begin{figure}[!b]
	\centering
	
	\vspace*{-0.15in}
	\includegraphics[width=3.4in]{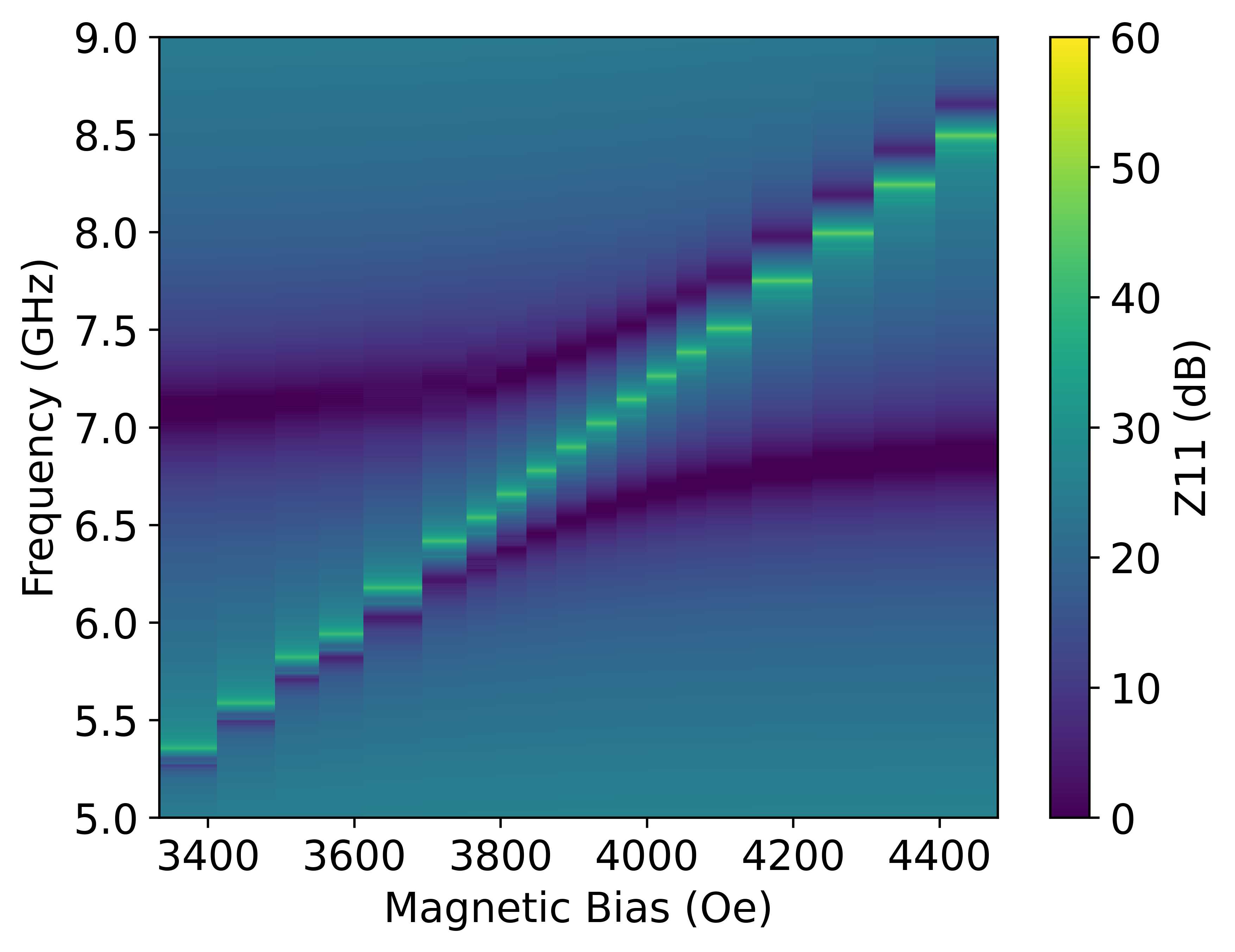}%
	
	\caption{Measured resonator impedance response at an input RF power of $\SI{-15}{\decibel m}$ as the applied magnetic field is swept . When the MSW is tuned to overlap with the distributed resonator, the modes interact strongly and an avoided crossing of the anti-resonance is clearly visible.}
	
	\label{frequency_splitting_model}
\end{figure}

We report a novel MSW resonator coupled to a distributed electromagnetic resonator using state-of-the-art micro-machining fabrication techniques. The magnetostatic coupling boost can be precisely controlled through the design of a radial stub termination and we demonstrate both $\SI{7}{\giga\hertz}$ and $\SI{18}{\giga\hertz}$ resonators for C-band and Ku-Band operation respectively. The resonator figure of merit ($k_{eff}^2 \cdot Q$) peaks at $50$ for the $\SI{7}{\giga\hertz}$ and at $121$ for the $\SI{18}{\giga\hertz}$ surpassing state-of-the-art piezoelectric resonators at similar frequencies \cite{barrera_thin-film_2023, yang_scaling_2018, fiagbenu_k-band_nodate, giribaldi_620_2023,kramer_thin-film_2023}. A sample filter topology utilizing these high figure of merit devices is discussed showing the design space trade-off between frequency tuning and bandwidth.
Interestingly, as the magnetic resonance is tuned to coincide with the distributed electromagnetic resonance a clear avoided crossing of the anti-resonance is observed in Fig. $\ref{frequency_splitting_model}$. The response is indicative of a hybridized mode where the magnetostatic wave (magnon) is coupled to the electromagnetic wave (photon). Substituting the gold transducers with a superconductor such as Nb or NbN and cooling the device to cryogenic temperatures, opens up applications in magnon-photon hybrid quantum systems. The lithographically defined devices allow flexible integration of the long-lifetime spin-waves in YIG with superconducting quantum platforms for novel quantum information processing, sensing, and networking opportunities \cite{xu_strong_2022, baity_strong_2021, xu_dynamical_2023}.

\section*{Data Availability}
The code and data used to produce the plots within this work will be released on the repository Zenodo upon publication.  \\ 

\vspace*{-0.15in}

\section*{Acknowledgments} 
Chip fabrication was performed at the Birck Nanotechnology Center at Purdue University. Resonator characterization and measurements were  was performed at Seng-Liang Wang Hall at Purdue.

	{\appendices
		\section*{Appendix}
              The resonator smith plot in Fig. $\ref{smith_plot}$ shows a higher-order spurious MSFVW width mode at the small loop near the normalized impedance, $Z=0.5+j1$. An ideal resonator should perfectly trace out the circumference of the smith chart (indicating a purely reactive response without loss) and be free of similar loops and notches indicating spurious resonances. The response of a MSFVW resonator cannot be free of spurious modes, but they can be further suppressed through resonator geometry optimization. The anti-resonance impedance (left side of the smith plot) can be improved by minimizing the resistance of the transducer. Significantly improving the MSFVW resonance impedance (right side of the smith plot) is more challenging and requires an investigation into the loss mechanisms of spin waves.
		\begin{figure}[H]
			\centering
			
			% \vspace*{-0.15in}
			\includegraphics[width=2.7in]{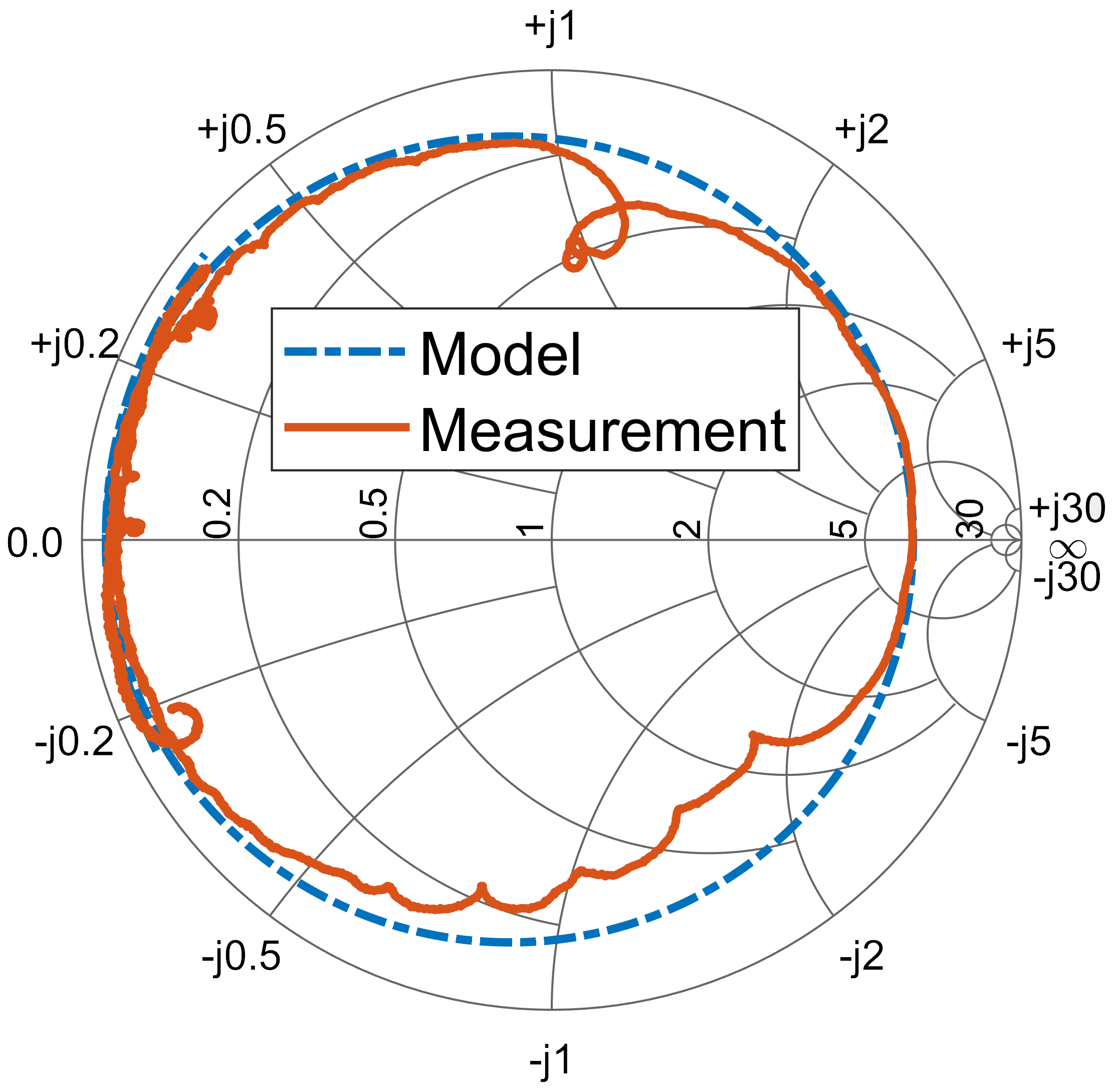}%
			
			\caption{Smith plot with $\SI{50}{\ohm}$ reference impedance of the measured resonator response (Fig. $\ref{Model_fitting}$) at $\SI{7714}{Oe}$ with $f_m=\SI{18.16}{\giga\hertz}$ alongside the fitted distributed circuit model response. }
			
			\label{smith_plot}
                % \vspace*{-0.1in}
		\end{figure}
	}

\balance

\bibliographystyle{IEEEtran}
\bibliography{MSFVW_Distributed_Resonator}

\begin{thebibliography}{10}
\providecommand{\url}[1]{#1}
\csname url@samestyle\endcsname
\providecommand{\newblock}{\relax}
\providecommand{\bibinfo}[2]{#2}
\providecommand{\BIBentrySTDinterwordspacing}{\spaceskip=0pt\relax}
\providecommand{\BIBentryALTinterwordstretchfactor}{4}
\providecommand{\BIBentryALTinterwordspacing}{\spaceskip=\fontdimen2\font plus
\BIBentryALTinterwordstretchfactor\fontdimen3\font minus \fontdimen4\font\relax}
\providecommand{\BIBforeignlanguage}[2]{{%
\expandafter\ifx\csname l@#1\endcsname\relax
\typeout{** WARNING: IEEEtran.bst: No hyphenation pattern has been}%
\typeout{** loaded for the language `#1'. Using the pattern for}%
\typeout{** the default language instead.}%
\else
\language=\csname l@#1\endcsname
\fi
#2}}
\providecommand{\BIBdecl}{\relax}
\BIBdecl
\renewcommand{\BIBentryALTinterwordstretchfactor}{4}

\bibitem{ariturk_element-level_2022}
\BIBentryALTinterwordspacing
G.~Ariturk, N.~R. Almuqati, and H.~H. Sigmarsson, ``\BIBforeignlanguage{en}{Element-{Level} {Microwave} {Filter} {Integration} in {Fully}-{Digital} {Phased} {Array} {Radar} {Systems}},'' in \emph{\BIBforeignlanguage{en}{2022 {IEEE} 22nd {Annual} {Wireless} and {Microwave} {Technology} {Conference} ({WAMICON})}}.\hskip 1em plus 0.5em minus 0.4em\relax Clearwater, FL, USA: IEEE, Apr. 2022, pp. 1--4. [Online]. Available: \url{https://ieeexplore.ieee.org/document/9786104/}
\BIBentrySTDinterwordspacing

\bibitem{cdi_proquest_reports_2564179163}
``\BIBforeignlanguage{eng}{Filtering out interference for next- generation wideband arrays},'' \emph{\BIBforeignlanguage{eng}{Microwave Journal}}, vol.~64, no.~8, pp. 35--36, 2021.

\bibitem{talisa_benefits_2016}
\BIBentryALTinterwordspacing
S.~H. Talisa \emph{et~al.}, ``\BIBforeignlanguage{en}{Benefits of {Digital} {Phased} {Array} {Radars}},'' \emph{\BIBforeignlanguage{en}{Proceedings of the IEEE}}, vol. 104, no.~3, pp. 530--543, Mar. 2016. [Online]. Available: \url{http://ieeexplore.ieee.org/document/7399690/}
\BIBentrySTDinterwordspacing

\bibitem{snyder_emerging_2021}
\BIBentryALTinterwordspacing
R.~V. Snyder \emph{et~al.}, ``\BIBforeignlanguage{en}{Emerging {Trends} in {Techniques} and {Technology} as {Applied} to {Filter} {Design}},'' \emph{\BIBforeignlanguage{en}{IEEE Journal of Microwaves}}, vol.~1, no.~1, pp. 317--344, Jan. 2021. [Online]. Available: \url{https://ieeexplore.ieee.org/document/9318743/}
\BIBentrySTDinterwordspacing

\bibitem{aigner_saw_2008}
\BIBentryALTinterwordspacing
R.~Aigner, ``\BIBforeignlanguage{en}{{SAW} and {BAW} technologies for {RF} filter applications: {A} review of the relative strengths and weaknesses},'' in \emph{\BIBforeignlanguage{en}{2008 {IEEE} {Ultrasonics} {Symposium}}}.\hskip 1em plus 0.5em minus 0.4em\relax Beijing, China: IEEE, Nov. 2008, pp. 582--589. [Online]. Available: \url{http://ieeexplore.ieee.org/document/4803350/}
\BIBentrySTDinterwordspacing

\bibitem{hagelauer_microwave_2023}
\BIBentryALTinterwordspacing
A.~Hagelauer \emph{et~al.}, ``\BIBforeignlanguage{en}{From {Microwave} {Acoustic} {Filters} to {Millimeter}-{Wave} {Operation} and {New} {Applications}},'' \emph{\BIBforeignlanguage{en}{IEEE Journal of Microwaves}}, vol.~3, no.~1, pp. 484--508, Jan. 2023. [Online]. Available: \url{https://ieeexplore.ieee.org/document/9994579/}
\BIBentrySTDinterwordspacing

\bibitem{yang_scaling_2018}
\BIBentryALTinterwordspacing
Y.~Yang, R.~Lu, and S.~Gong, ``\BIBforeignlanguage{en}{Scaling {Acoustic} {Filters} {Towards} {5G}},'' in \emph{\BIBforeignlanguage{en}{2018 {IEEE} {International} {Electron} {Devices} {Meeting} ({IEDM})}}.\hskip 1em plus 0.5em minus 0.4em\relax San Francisco, CA: IEEE, Dec. 2018, pp. 39.6.1--39.6.4. [Online]. Available: \url{https://ieeexplore.ieee.org/document/8614699/}
\BIBentrySTDinterwordspacing

\bibitem{barrera_thin-film_2023}
O.~Barrera \emph{et~al.}, ``Thin-{Film} {Lithium} {Niobate} {Acoustic} {Filter} at 23.5 {GHz} with 2.38 {dB} {IL} and 18.2\% {FBW},'' \emph{arXiv preprint arXiv:2307.04559}, 2023.

\bibitem{fiagbenu_k-band_nodate}
Izhar \emph{et~al.}, ``A k-band bulk acoustic wave resonator using periodically poled al0.72sc0.28n,'' \emph{IEEE Electron Device Letters}, vol.~44, no.~7, pp. 1196--1199, 2023.

\bibitem{zou_aluminum_2022}
\BIBentryALTinterwordspacing
Y.~Zou \emph{et~al.}, ``Aluminum scandium nitride thin-film bulk acoustic resonators for {5G} wideband applications,'' \emph{Microsystems \& Nanoengineering}, vol.~8, no.~1, p. 124, Nov. 2022. [Online]. Available: \url{https://doi.org/10.1038/s41378-022-00457-0}
\BIBentrySTDinterwordspacing

\bibitem{giribaldi_compact_2023}
\BIBentryALTinterwordspacing
G.~Giribaldi \emph{et~al.}, ``\BIBforeignlanguage{en}{Compact and wideband nanoacoustic pass-band filters for future {5G} and {6G} cellular radios},'' In Review, preprint, Feb. 2023. [Online]. Available: \url{https://doi.org/10.21203/rs.3.rs-2569732/v1}
\BIBentrySTDinterwordspacing

\bibitem{adam_msw_1985}
\BIBentryALTinterwordspacing
J.~Adam, ``An {MSW} {Tunable} {Bandpass} {Filter},'' in \emph{{IEEE} 1985 {Ultrasonics} {Symposium}}.\hskip 1em plus 0.5em minus 0.4em\relax San Francisco, CA, USA: IEEE, 1985, pp. 157--162. [Online]. Available: \url{http://ieeexplore.ieee.org/document/1535436/}
\BIBentrySTDinterwordspacing

\bibitem{hanna_single_1988}
\BIBentryALTinterwordspacing
S.~Hanna and S.~Zeroug, ``\BIBforeignlanguage{en}{Single and coupled {MSW} resonators for microwave channelizers},'' \emph{\BIBforeignlanguage{en}{IEEE Transactions on Magnetics}}, vol.~24, no.~6, pp. 2808--2810, Nov. 1988. [Online]. Available: \url{http://ieeexplore.ieee.org/document/92252/}
\BIBentrySTDinterwordspacing

\bibitem{yang_low-loss_2013}
\BIBentryALTinterwordspacing
G.-M. Yang \emph{et~al.}, ``\BIBforeignlanguage{en}{Low-{Loss} {Magnetically} {Tunable} {Bandpass} {Filters} {With} {YIG} {Films}},'' \emph{\BIBforeignlanguage{en}{IEEE Transactions on Magnetics}}, vol.~49, no.~9, pp. 5063--5068, Sep. 2013. [Online]. Available: \url{http://ieeexplore.ieee.org/document/6508942/}
\BIBentrySTDinterwordspacing

\bibitem{zhang_nonreciprocal_2020}
\BIBentryALTinterwordspacing
Y.~Zhang \emph{et~al.}, ``\BIBforeignlanguage{en}{Nonreciprocal {Isolating} {Bandpass} {Filter} {With} {Enhanced} {Isolation} {Using} {Metallized} {Ferrite}},'' \emph{\BIBforeignlanguage{en}{IEEE Transactions on Microwave Theory and Techniques}}, vol.~68, no.~12, pp. 5307--5316, Dec. 2020. [Online]. Available: \url{https://ieeexplore.ieee.org/document/9238458/}
\BIBentrySTDinterwordspacing

\bibitem{noauthor_yig_nodate}
\BIBentryALTinterwordspacing
``{YIG} {Filters}.'' [Online]. Available: \url{https://www.teledynedefenseelectronics.com/wireless/YIG%20Products/Pages/YIG%20Filters.aspx}
\BIBentrySTDinterwordspacing

\bibitem{fletcher_ferrimagnetic_1959}
\BIBentryALTinterwordspacing
P.~C. Fletcher and R.~O. Bell, ``\BIBforeignlanguage{en}{Ferrimagnetic {Resonance} {Modes} in {Spheres}},'' \emph{\BIBforeignlanguage{en}{Journal of Applied Physics}}, vol.~30, no.~5, pp. 687--698, May 1959. [Online]. Available: \url{https://pubs.aip.org/jap/article/30/5/687/162486/Ferrimagnetic-Resonance-Modes-in-Spheres}
\BIBentrySTDinterwordspacing

\bibitem{carter_magnetically-tunable_1961}
\BIBentryALTinterwordspacing
P.~Carter, ``\BIBforeignlanguage{en}{Magnetically-{Tunable} {Microwave} {Filters} {Using} {Single}-{Crystal} {Yttrium}-{Iron}-{Garnet} {Resonators}},'' \emph{\BIBforeignlanguage{en}{IEEE Transactions on Microwave Theory and Techniques}}, vol.~9, no.~3, pp. 252--260, May 1961. [Online]. Available: \url{http://ieeexplore.ieee.org/document/1125316/}
\BIBentrySTDinterwordspacing

\bibitem{ishak_magnetostatic_1988}
\BIBentryALTinterwordspacing
W.~Ishak, ``\BIBforeignlanguage{en}{Magnetostatic wave technology: a review},'' \emph{\BIBforeignlanguage{en}{Proceedings of the IEEE}}, vol.~76, no.~2, pp. 171--187, Feb. 1988. [Online]. Available: \url{http://ieeexplore.ieee.org/document/4393/}
\BIBentrySTDinterwordspacing

\bibitem{du_frequency_2023}
\BIBentryALTinterwordspacing
X.~Du \emph{et~al.}, ``Frequency tunable magnetostatic wave filters with zero static power magnetic biasing circuitry,'' no. {arXiv}:2308.00907. [Online]. Available: \url{http://arxiv.org/abs/2308.00907}
\BIBentrySTDinterwordspacing

\bibitem{tikhonov_temperature_2013}
\BIBentryALTinterwordspacing
V.~V. Tikhonov \emph{et~al.}, ``\BIBforeignlanguage{en}{Temperature stabilization of spin-wave ferrite devices},'' \emph{\BIBforeignlanguage{en}{Journal of Communications Technology and Electronics}}, vol.~58, no.~1, pp. 75--81, Jan. 2013. [Online]. Available: \url{http://link.springer.com/10.1134/S1064226913010075}
\BIBentrySTDinterwordspacing

\bibitem{devitt}
\BIBentryALTinterwordspacing
C.~Devitt \emph{et~al.}, ``\BIBforeignlanguage{en}{An {Edge}-{Coupled} {Magnetostatic} {Bandpass} {Filter}},'' no. {arXiv}:2312.10583, Dec. 2023. [Online]. Available: \url{https://arxiv.org/abs/2312.10583}
\BIBentrySTDinterwordspacing

\bibitem{dai_octave-tunable_2020}
\BIBentryALTinterwordspacing
S.~Dai, S.~A. Bhave, and R.~Wang, ``\BIBforeignlanguage{en}{Octave-{Tunable} {Magnetostatic} {Wave} {YIG} {Resonators} on a {Chip}},'' \emph{\BIBforeignlanguage{en}{IEEE Transactions on Ultrasonics, Ferroelectrics, and Frequency Control}}, vol.~67, no.~11, pp. 2454--2460, Nov. 2020. [Online]. Available: \url{https://ieeexplore.ieee.org/document/9108306/}
\BIBentrySTDinterwordspacing

\bibitem{marcelli_band-pass_2004}
\BIBentryALTinterwordspacing
R.~Marcelli, G.~Sajin, and A.~Cismaru, ``\BIBforeignlanguage{en}{Band-pass magnetostatic wave resonators on micromachined silicon substrate},'' \emph{\BIBforeignlanguage{en}{Review of Scientific Instruments}}, vol.~75, no.~4, pp. 1127--1133, Apr. 2004. [Online]. Available: \url{https://pubs.aip.org/rsi/article/75/4/1127/466503/Band-pass-magnetostatic-wave-resonators-on}
\BIBentrySTDinterwordspacing

\bibitem{costa_compact_2021}
\BIBentryALTinterwordspacing
J.~D. Costa \emph{et~al.}, ``\BIBforeignlanguage{en}{Compact tunable {YIG}-based {RF} resonators},'' \emph{\BIBforeignlanguage{en}{Applied Physics Letters}}, vol. 118, no.~16, p. 162406, Apr. 2021. [Online]. Available: \url{https://pubs.aip.org/apl/article/118/16/162406/1062227/Compact-tunable-YIG-based-RF-resonators}
\BIBentrySTDinterwordspacing

\bibitem{connelly2023principles}
D.~A. Connelly, ``\BIBforeignlanguage{English}{Principles, modeling, and measurement towards efficient microwave spin-wave circuits},'' Ph.D. dissertation, 2023.

\bibitem{feng_micromachined_2023}
\BIBentryALTinterwordspacing
Y.~Feng \emph{et~al.}, ``\BIBforeignlanguage{en}{Micromachined {Tunable} {Magnetostatic} {Forward} {Volume} {Wave} {Bandstop} {Filter}},'' \emph{\BIBforeignlanguage{en}{IEEE Microwave and Wireless Technology Letters}}, vol.~33, no.~6, pp. 807--810, Jun. 2023. [Online]. Available: \url{https://ieeexplore.ieee.org/document/10111074/}
\BIBentrySTDinterwordspacing

\bibitem{gao_design_2022}
\BIBentryALTinterwordspacing
Q.~Gao \emph{et~al.}, ``\BIBforeignlanguage{en}{Design {RF} {Magnetic} {Devices} {With} {Linear} and {Nonlinear} {Equivalent} {Circuit} {Models}: {Demystify} {RF} {Magnetics} {With} {Equivalent} {Circuit} {Models}},'' \emph{\BIBforeignlanguage{en}{IEEE Microwave Magazine}}, vol.~23, no.~11, pp. 28--47, Nov. 2022. [Online]. Available: \url{https://ieeexplore.ieee.org/document/9910195/}
\BIBentrySTDinterwordspacing

\bibitem{gao_equivalent_2021}
\BIBentryALTinterwordspacing
Q.~Gao, ``\BIBforeignlanguage{English}{An {Equivalent} {Circuit} {Model} for {Tunable} {Bandpass} {Filters} {Based} on {Ferromagnetic} {Resonance}},'' Master's thesis, University of California, Los Angeles, United States -- California, 2021. [Online]. Available: \url{https://www.proquest.com/dissertations-theses/equivalent-circuit-model-tunable-bandpass-filters/docview/2555310965/se-2?accountid=13360}
\BIBentrySTDinterwordspacing

\bibitem{cui_coupling_2019}
\BIBentryALTinterwordspacing
H.~Cui, Z.~Yao, and Y.~E. Wang, ``\BIBforeignlanguage{en}{Coupling {Electromagnetic} {Waves} to {Spin} {Waves}: {A} {Physics}-{Based} {Nonlinear} {Circuit} {Model} for {Frequency}-{Selective} {Limiters}},'' \emph{\BIBforeignlanguage{en}{IEEE Transactions on Microwave Theory and Techniques}}, vol.~67, no.~8, pp. 3221--3229, Aug. 2019. [Online]. Available: \url{https://ieeexplore.ieee.org/document/8736517/}
\BIBentrySTDinterwordspacing

\bibitem{stancil_theory_1993}
\BIBentryALTinterwordspacing
D.~D. Stancil, \emph{\BIBforeignlanguage{en}{Theory of {Magnetostatic} {Waves}}}, New York, NY, 1993. [Online]. Available: \url{http://link.springer.com/10.1007/978-1-4613-9338-2}
\BIBentrySTDinterwordspacing

\bibitem{ishak_tunable_1986}
\BIBentryALTinterwordspacing
W.~Ishak and {Kok-Wai Chang}, ``\BIBforeignlanguage{en}{Tunable {Microwave} {Resonators} {Using} {Magnetostatic} {Wave} in {YIG} {Films}},'' \emph{\BIBforeignlanguage{en}{IEEE Transactions on Microwave Theory and Techniques}}, vol.~34, no.~12, pp. 1383--1393, Dec. 1986. [Online]. Available: \url{http://ieeexplore.ieee.org/document/1133553/}
\BIBentrySTDinterwordspacing

\bibitem{ishak_tunable_1988}
\BIBentryALTinterwordspacing
W.~Ishak \emph{et~al.}, ``\BIBforeignlanguage{en}{Tunable microwave resonators and oscillators using magnetostatic waves},'' \emph{\BIBforeignlanguage{en}{IEEE Transactions on Ultrasonics, Ferroelectrics and Frequency Control}}, vol.~35, no.~3, pp. 396--405, May 1988. [Online]. Available: \url{http://ieeexplore.ieee.org/document/20461/}
\BIBentrySTDinterwordspacing

\bibitem{larson_modified_2000}
\BIBentryALTinterwordspacing
J.~Larson \emph{et~al.}, ``\BIBforeignlanguage{en}{Modified {Butterworth}-{Van} {Dyke} circuit for {FBAR} resonators and automated measurement system},'' in \emph{\BIBforeignlanguage{en}{2000 {IEEE} {Ultrasonics} {Symposium}. {Proceedings}. {An} {International} {Symposium} ({Cat}. {No}.{00CH37121})}}.\hskip 1em plus 0.5em minus 0.4em\relax San Juan, Puerto Rico, USA: IEEE, 2000, pp. 863--868 vol.1. [Online]. Available: \url{https://ieeexplore.ieee.org/document/922679/}
\BIBentrySTDinterwordspacing

\bibitem{psychogiou_hybrid_2015}
\BIBentryALTinterwordspacing
D.~Psychogiou \emph{et~al.}, ``\BIBforeignlanguage{en}{Hybrid {Acoustic}-{Wave}-{Lumped}-{Element} {Resonators} ({AWLRs}) for {High-Q} {Bandpass} {Filters} {With} {Quasi}-{Elliptic} {Frequency} {Response}},'' \emph{\BIBforeignlanguage{en}{IEEE Transactions on Microwave Theory and Techniques}}, vol.~63, no.~7, pp. 2233--2244, Jul. 2015. [Online]. Available: \url{http://ieeexplore.ieee.org/document/7122936/}
\BIBentrySTDinterwordspacing

\bibitem{yang_x-band_2021}
\BIBentryALTinterwordspacing
Y.~Yang, L.~Gao, and S.~Gong, ``\BIBforeignlanguage{en}{X-{Band} {Miniature} {Filters} {Using} {Lithium} {Niobate} {Acoustic} {Resonators} and {Bandwidth} {Widening} {Technique}},'' \emph{\BIBforeignlanguage{en}{IEEE Transactions on Microwave Theory and Techniques}}, vol.~69, no.~3, pp. 1602--1610, Mar. 2021. [Online]. Available: \url{https://ieeexplore.ieee.org/document/9337207/}
\BIBentrySTDinterwordspacing

\bibitem{truitt_efficient_2007}
\BIBentryALTinterwordspacing
P.~A. Truitt \emph{et~al.}, ``\BIBforeignlanguage{en}{Efficient and {Sensitive} {Capacitive} {Readout} of {Nanomechanical} {Resonator} {Arrays}},'' \emph{\BIBforeignlanguage{en}{Nano Letters}}, vol.~7, no.~1, pp. 120--126, Jan. 2007. [Online]. Available: \url{https://pubs.acs.org/doi/10.1021/nl062278g}
\BIBentrySTDinterwordspacing

\bibitem{cagdaser_low-voltage_2012}
\BIBentryALTinterwordspacing
B.~Cagdaser and B.~E. Boser, ``\BIBforeignlanguage{en}{Low-{Voltage} {Electrostatic} {Actuation} {With} {Inherent} {Position} {Feedback}},'' \emph{\BIBforeignlanguage{en}{Journal of Microelectromechanical Systems}}, vol.~21, no.~5, pp. 1187--1196, Oct. 2012. [Online]. Available: \url{http://ieeexplore.ieee.org/document/6199945/}
\BIBentrySTDinterwordspacing

\bibitem{barzanjeh_mechanical_2017}
\BIBentryALTinterwordspacing
S.~Barzanjeh \emph{et~al.}, ``\BIBforeignlanguage{en}{Mechanical on-chip microwave circulator},'' \emph{\BIBforeignlanguage{en}{Nature Communications}}, vol.~8, no.~1, p. 953, Oct. 2017. [Online]. Available: \url{https://www.nature.com/articles/s41467-017-01304-x}
\BIBentrySTDinterwordspacing

\bibitem{yang_modified_2016}
\BIBentryALTinterwordspacing
Q.~Yang \emph{et~al.}, ``\BIBforeignlanguage{en}{A {Modified} {Lattice} {Configuration} {Design} for {Compact} {Wideband} {Bulk} {Acoustic} {Wave} {Filter} {Applications}},'' \emph{\BIBforeignlanguage{en}{Micromachines}}, vol.~7, no.~8, p. 133, Aug. 2016. [Online]. Available: \url{http://www.mdpi.com/2072-666X/7/8/133}
\BIBentrySTDinterwordspacing

\bibitem{xie_sub-terahertz_2023}
\BIBentryALTinterwordspacing
J.~Xie \emph{et~al.}, ``\BIBforeignlanguage{en}{Sub-terahertz electromechanics},'' \emph{\BIBforeignlanguage{en}{Nature Electronics}}, vol.~6, no.~4, pp. 301--306, Mar. 2023. [Online]. Available: \url{https://www.nature.com/articles/s41928-023-00942-y}
\BIBentrySTDinterwordspacing

\bibitem{wadell_transmission_1991}
B.~C. Wadell, \emph{\BIBforeignlanguage{eng}{Transmission line design handbook}}, ser. The {Artech} {House} microwave library.\hskip 1em plus 0.5em minus 0.4em\relax Boston: Artech House, 1991, publication Title: Transmission line design handbook.

\bibitem{asao_tunable_1995}
H.~Asao \emph{et~al.}, ``A tunable oscillator using magnetostatic forward-volume wave resonator with wide strip transducer,'' \emph{Electronics and Communications in Japan (Part II: Electronics)}, vol.~78, no.~5, pp. 81--91, 1995, publisher: Wiley Online Library.

\bibitem{zhang_process_2004}
\BIBentryALTinterwordspacing
S.~Zhang \emph{et~al.}, ``\BIBforeignlanguage{en}{Process for uniformly electroplating a patterned wafer with electrically isolated devices},'' \emph{\BIBforeignlanguage{en}{Journal of Vacuum Science \& Technology A: Vacuum, Surfaces, and Films}}, vol.~22, no.~3, pp. 1079--1082, May 2004. [Online]. Available: \url{https://pubs.aip.org/jva/article/22/3/1079/1072216/Process-for-uniformly-electroplating-a-patterned}
\BIBentrySTDinterwordspacing

\bibitem{pulskamp_monolithically_2009}
\BIBentryALTinterwordspacing
J.~Pulskamp \emph{et~al.}, ``\BIBforeignlanguage{en}{Monolithically {Integrated} {Piezomems} {SP2T} {Switch} and {Contour}-{Mode} {Filters}},'' in \emph{\BIBforeignlanguage{en}{2009 {IEEE} 22nd {International} {Conference} on {Micro} {Electro} {Mechanical} {Systems}}}.\hskip 1em plus 0.5em minus 0.4em\relax Sorrento, Italy: IEEE, Jan. 2009, pp. 900--903. [Online]. Available: \url{http://ieeexplore.ieee.org/document/4805529/}
\BIBentrySTDinterwordspacing

\bibitem{pozar_microwave_2012}
D.~M. Pozar, \emph{\BIBforeignlanguage{eng}{Microwave engineering}}, 4th~ed.\hskip 1em plus 0.5em minus 0.4em\relax Hoboken, NJ: Wiley, 2012, publication Title: Microwave engineering.

\bibitem{swanson_narrow-band_2007}
\BIBentryALTinterwordspacing
D.~Swanson, ``\BIBforeignlanguage{en}{Narrow-band microwave filter design},'' \emph{\BIBforeignlanguage{en}{IEEE Microwave Magazine}}, vol.~8, no.~5, pp. 105--114, Oct. 2007. [Online]. Available: \url{http://ieeexplore.ieee.org/document/4383441/}
\BIBentrySTDinterwordspacing

\bibitem{dishal_alignment_1951}
\BIBentryALTinterwordspacing
M.~Dishal, ``\BIBforeignlanguage{en}{Alignment and {Adjustment} of {Synchronously} {Tuned} {Multiple}-{Resonant}-{Circuit} {Filters}},'' \emph{\BIBforeignlanguage{en}{Proceedings of the IRE}}, vol.~39, no.~11, pp. 1448--1455, Nov. 1951. [Online]. Available: \url{http://ieeexplore.ieee.org/document/4050569/}
\BIBentrySTDinterwordspacing

\bibitem{giribaldi_620_2023}
\BIBentryALTinterwordspacing
G.~Giribaldi, L.~Colombo, and M.~Rinaldi, ``\BIBforeignlanguage{en}{6–20 {GHz} 30\% {ScAlN} {Lateral} {Field}-{Excited} {Cross}-{Sectional} {Lamé} {Mode} {Resonators} for {Future} {Mobile} {RF} {Front} {Ends}},'' \emph{\BIBforeignlanguage{en}{IEEE Transactions on Ultrasonics, Ferroelectrics, and Frequency Control}}, vol.~70, no.~10, pp. 1201--1212, Oct. 2023. [Online]. Available: \url{https://ieeexplore.ieee.org/document/10243055/}
\BIBentrySTDinterwordspacing

\bibitem{kramer_thin-film_2023}
\BIBentryALTinterwordspacing
J.~Kramer \emph{et~al.}, ``\BIBforeignlanguage{en}{Thin-{Film} {Lithium} {Niobate} {Acoustic} {Resonator} with {High} {Q} of 237 and k $^{\textrm{2}}$ of 5.1\% at 50.74 {GHz}},'' in \emph{\BIBforeignlanguage{en}{2023 {Joint} {Conference} of the {European} {Frequency} and {Time} {Forum} and {IEEE} {International} {Frequency} {Control} {Symposium} ({EFTF}/{IFCS})}}.\hskip 1em plus 0.5em minus 0.4em\relax Toyama, Japan: IEEE, May 2023, pp. 1--4. [Online]. Available: \url{https://ieeexplore.ieee.org/document/10272149/}
\BIBentrySTDinterwordspacing

\bibitem{xu_strong_2022}
\BIBentryALTinterwordspacing
Q.~Xu \emph{et~al.}, ``\BIBforeignlanguage{en}{Strong photon-magnon coupling using a lithographically defined organic ferrimagnet},'' Dec. 2022, arXiv:2212.04423 [cond-mat, physics:quant-ph]. [Online]. Available: \url{http://arxiv.org/abs/2212.04423}
\BIBentrySTDinterwordspacing

\bibitem{baity_strong_2021}
\BIBentryALTinterwordspacing
P.~G. Baity \emph{et~al.}, ``\BIBforeignlanguage{en}{Strong magnon–photon coupling with chip-integrated {YIG} in the zero-temperature limit},'' \emph{\BIBforeignlanguage{en}{Applied Physics Letters}}, vol. 119, no.~3, p. 033502, Jul. 2021. [Online]. Available: \url{https://pubs.aip.org/apl/article/119/3/033502/41793/Strong-magnon-photon-coupling-with-chip-integrated}
\BIBentrySTDinterwordspacing

\bibitem{xu_dynamical_2023}
\BIBentryALTinterwordspacing
J.~Xu \emph{et~al.}, ``\BIBforeignlanguage{en}{Dynamical control in hybrid magnonics},'' in \emph{\BIBforeignlanguage{en}{Spintronics {XVI}}}, J.-E. Wegrowe, M.~Razeghi, and J.~S. Friedman, Eds.\hskip 1em plus 0.5em minus 0.4em\relax San Diego, United States: SPIE, Sep. 2023, p.~74. [Online]. Available: \url{https://www.spiedigitallibrary.org/conference-proceedings-of-spie/12656/2677206/Dynamical-control-in-hybrid-magnonics/10.1117/12.2677206.full}
\BIBentrySTDinterwordspacing

\end{thebibliography}

\vfill

\end{document}